\documentclass[12pt,twocolumn,twoside]{article}

\usepackage{titlesec}
\usepackage{fancyhdr}
\usepackage{authblk}
\usepackage{siunitx}
\usepackage{setspace,verbatim}
\usepackage{transparent}

\usepackage{xcolor,colortbl}
\usepackage{graphicx}
\usepackage{longtable}
\usepackage{booktabs}
\usepackage{multirow}
\usepackage[colorlinks, linkcolor=blue, anchorcolor=blue, citecolor=blue]{hyperref}
\usepackage{subcaption}
\usepackage{amsfonts}
\usepackage{amssymb}
\usepackage{amsmath}
\usepackage{lscape}
\usepackage{blindtext}
\usepackage{subcaption}
\usepackage{textcomp}
\usepackage{gensymb}
\usepackage{threeparttable}
\usepackage{booktabs, caption, makecell}
\usepackage{url}

\usepackage{breakurl}
\usepackage{hyperref}

\usepackage[top=1in,bottom=1in,left=1in,right=1in]{geometry}

\def\01{\ensuremath{0\mathord{-}1}}
\def\st{\mathop{\rm s.t.}}

\usepackage{algorithm}
\usepackage{algorithmicx}
\usepackage{algpseudocode}

\usepackage[customcolors]{hf-tikz}

\tikzset{style green/.style={
    set fill color=green!50!lime!60,
    set border color=white,
  },
  style cyan/.style={
    set fill color=cyan!90!blue!60,
    set border color=white,
  },
  style orange/.style={
    set fill color=orange!80!red!60,
    set border color=white,
  },
  hor/.style={
    above left offset={-0.15,0.31},
    below right offset={0.15,-0.125},
    #1
  },
  ver/.style={
    above left offset={-0.1,0.3},
    below right offset={0.15,-0.15},
    #1
  }
}

\titleformat*{\section}{\normalsize\bfseries\sffamily}
\titleformat*{\subsection}{\small\bfseries\sffamily}
\titleformat*{\subsubsection}{\small\bfseries\sffamily}

\usepackage[font=small,labelfont=bf]{caption}

\newcommand\shorttitle{Domain reduction techniques for global NLP and MINLP optimization}
\newcommand\authors{Puranik and Sahinidis}

\fancypagestyle{firststyle}
{
   \fancyhf{}

   \fancyfoot[L]{\footnotesize\scshape \copyright 2017. This manuscript version is made available under the CC-BY-NC-ND 4.0 license \url{http://creativecommons.org/licenses/by-nc-nd/4.0/}.  The formal publication of this article is available at \url{https://link.springer.com/article/10.1007/s10601-016-9267-5}.}
}

\title{\textbf{Domain reduction techniques for global NLP and MINLP optimization}}
\author[1]{Y. Puranik}
\author[1]{Nikolaos~V.~Sahinidis}
\affil[1]{Department of Chemical Engineering, Carnegie Mellon University, Pittsburgh, PA 15213, USA}
\date{}

\begin{document}

\begin{singlespacing}
\twocolumn[{%
%%% title
\maketitle
\thispagestyle{firststyle}

% Define bibstyle
\bibliographystyle{plainnat}

%\doublespacing
%\baselinestretch

%\section**{Abstract}
%\vspace{-5pt}
%\pagenumbering{gobble}
\footnotesize
\vspace{-1cm}
\begin{abstract}
Optimization solvers routinely utilize presolve techniques, including model simplification, reformulation and domain reduction techniques. Domain reduction techniques are especially important in speeding up convergence to the global optimum for challenging nonconvex nonlinear programming (NLP) and mixed-integer nonlinear programming (MINLP) optimization problems. In this work, we survey the various techniques used for domain reduction of NLP and MINLP optimization problems. We also present a computational analysis of the impact of these techniques on the performance of various widely available global solvers on a collection of 1740 test problems.
\end{abstract}


\vspace*{0.8cm}

}]

\footnotesize
%\newpage

\section{Introduction}
\label{sec:intro}

We consider the following mixed-integer nonlinear programming problem (MINLP):
\begin{equation} \label{eq:prob}
\left. \begin{array}{lll}
\min  \;\;& f(\vec{x}) \\
\st & \vec{g(\vec{x})} \leq 0 \\
 & \vec{x}_{l}\leq \vec{x} \leq \vec{x}_{u} \\
 & \vec{x} \in \mathbb{R}^{n-m}\times \mathbb{Z}^m
\end{array} \right\}
\end{equation}
MINLP is a very general representation for optimization problems and includes linear programming (LP), mixed-integer linear programming (MIP) and nonlinear programming (NLP) in its subclasses. A variety of applications in diverse fields are routinely formulated using this framework including water network design~\cite{j10,fb12}, hydro energy systems management~\cite{cpm11}, protein folding~\cite{n97-pro}, robust control~\cite{bb92}, trim loss~\cite{hwpk98}, heat exchanger network synthesis~\cite{fs02}, gas networks~\cite{mmm06}, transportation~\cite{fhsv10}, chemical process operations~\cite{g05}, chemical process synthesis~\cite{gcy00}, crystallographic imaging~\cite{ss09}, and seizure predictions~\cite{pci04}. Modelling via nonconvex objective functions or constraints is necessitated for many of these practical applications. Aside from the combinatorial complexity introduced by the integer variables, nonconvexities in objective function or the feasible region lead to multiple local minima and provide a challenge to the optimization of such problems.

Branch-and-bound~\cite{mjss16} based methods can be exploited to solve Problem~\ref{eq:prob} to global optimality. Inspired by branch-and-bound for discrete programming problems~\cite{ld60}, branch-and-bound was adapted for continuous problems by Falk and Soland~\cite{fs69}. The algorithm proceeds by bounding the global optimum by a valid lower and upper bound throughout the search process. Whenever these bounds are within an acceptable tolerance, the algorithm terminates with the upper bound as a (near-)global optimum. The algorithm exhaustively searches the space by branching on variables to divide the space into subdomains. The lower and upper bounding procedures are recursively applied to the new subdomains in a tree search until the bounds converge. Branch-and-bound algorithms are known to terminate within $\epsilon$-accuracy ($\epsilon>0$) to a global optimum as long as branching, node selection and lower bounding are designed to satisfy certain conditions~\cite{ht96}. For special cases, the true global optimum can be finitely achieved with branch-and-bound~\cite{ss98,ks00,bv08}. The success of branch-and-bound methods for global optimization is evident from the numerous software implementations available, including ANTIGONE~\cite{mf:14:anti}, BARON~\cite{ts:comp:04}, Couenne~\cite{b+:09:oms},  LindoGlobal~\cite{ls:09:oms} and SCIP~\cite{a09:scip}.

In this work, we survey the various domain reduction techniques that are employed within branch-and-bound algorithms. While these techniques are not necessary to ensure convergence to the global optimum, they typically speed up convergence. These techniques often exploit feasibility analysis to eliminate infeasible parts of the search space. Alternatively, the methods can also utilize optimality arguments to shrink the search space while ensuring at least one optimal solution is retained. Domain reduction techniques constitute the major component of the solution methods for satisfiability problems through unit propagation~\cite{ms99} and for constraint programming (CP) through various filtering algorithms that achieve differing levels of consistencies~\cite{b06}. They are also exploited in artificial intelligence (AI)~\cite{d87} and interval analysis~\cite{n90,k96}. Some of the other names used in the literature for these methods include bound propagation, bounds tightening, bound strengthening, domain filtering, bound reduction and range reduction.

\begin{figure}[htbp]
\centering
\includegraphics[scale = 0.45]{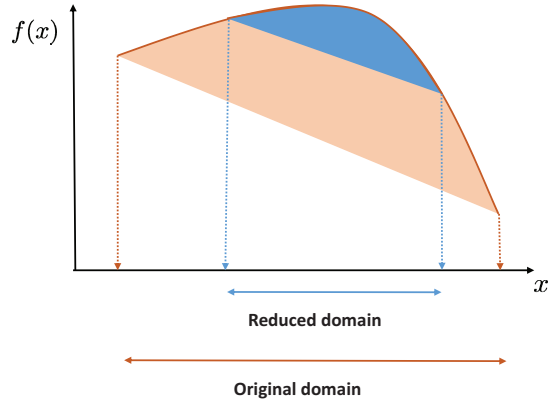}
\caption{Reduction in domains leads to tighter relaxations}
\label{fig:tight}
\end{figure}

Mathematical-programming-based methods for solving nonconvex MINLPs often rely on the solution of a relaxation problem for finding a valid lower bound. The strength of the relaxations employed for lower bounding depends on the diameter of the feasible region. Smaller domains lead to tighter relaxations. For example, Figure~\ref{fig:tight} shows the convex relaxation for a simple univariate concave function. A convex relaxation for a given function defined on a nonempty convex set is a convex function that underestimates the given function on its domain. The convex relaxation over the reduced domain provides a better approximation for the univariate concave function thereby providing better lower bounds. Domain reduction techniques not only reduce the diameter of the search space, but they also improve the tightness of convex relaxations.

Domain reduction techniques have been extensively studied in various communities including AI, CP, mathematical programming, interval analysis and computer science where they can be viewed as chaotic iterations~\cite{a99}. The primary objective of this paper is to review key domain reduction techniques applicable to (nonconvex) NLPs and MINLPs and point to connections between methods from constraint programming and interval arithmetic communities wherever applicable. We also present results on standard test libraries with different global solvers to demonstrate the impact of domain reduction strategies on their performance.

The remainder of the paper is organized as follows. Representation of general MINLPs through factorable reformulations and directed acyclic graphs is described in Section~\ref{sec:repr}. Methods for constraint propagation and bounds tightening in global optimization of MINLPs often rely on interval arithmetic. A brief introduction to this topic is provided in Section~\ref{sec:int:ar}. We introduce presolving for optimization in Section~\ref{sec:presolve}. Domain reduction techniques that rely on eliminating infeasible regions for the problem are described in Section~\ref{sec:fbbt}. Techniques that utilize optimality arguments for carrying out domain reduction are described in Section~\ref{sec:subopt}. Computational tradeoffs in the implementation of some of the advanced techniques are discussed in Section~\ref{sec:strat}. The computational impact of many of these techniques on the performance of widely available solvers is investigated in Section~\ref{sec:comp}. Finally, we conclude in Section~\ref{conclusions}.

\section{Representation}
\label{sec:repr}
A crucial step in the branch-and-bound algorithm is the construction of relaxations. One of the most widely used methods for this purpose is the idea of factorable reformulations. It involves splitting a problem into basic atomic functions that are utilized for computing the function, an idea exploited by McCormick~\cite{mc76}, who developed a technique that constructed non-differentiable relaxations of optimization problems. Ryoo and Sahinidis~\cite{rs96} gain differentiability by introducing new variables and equations for each of the intermediate functional forms. These functional forms are simple in nature like the bilinear term. A similar idea was proposed by Kearfott~\cite{k91} where new variables and equations are introduced to decompose nonlinearities to allow for more accurate computations of interval Jacobian matrices.

Consider the following example:
\begin{equation} \label{ex}
\left. \begin{array}{lll}
\min \; \;& 3x + 4y \\
& x - y \leq 4 \\
&xy \leq 3 \\
& x^2 + y^2 \geq 1 \\
& 1 \leq x \leq 5 \\
& 1 \leq y \leq 5 \\
\end{array} \right\}
\end{equation}
A factorable reformulation can proceed by introducing a new variable for every nonlinearity occurring in the model. We replace $z_1$ for $x^2$, $z_2$ for $y^2$ and $z_3$ for $xy$.

A factorable reformulation of the model is thus given by:
\begin{align}
\min \quad & 3x + 4y  \notag \\
& x - y \leq 4\notag \\
&z_3 \leq 3 \label{eq:non1} \\
& z_1 + z_2 \geq 1 \notag\\
& z_1 = x^2 \notag \\
& z_2 = y^2 \notag \\
& z_3 = xy \label{eq:non2} \\
& 1 \leq x \leq 5 \notag\\
& 1 \leq y \leq 5 \notag
\end{align}
It is trivial to outer approximate the univariate terms of this model (cf. Figure~\ref{fig:tight}), while the bilinear term in \ref{eq:non2} may be outer approximated by its convex and concave envelopes~\cite{mc76}.  The combination of these outer approximators provides a convex relaxation of the original problem.  In general, factorable reformulations decompose the problem into simpler functional forms for which convex relaxations are known. Thus, a convex relaxation can be obtained by reformulating a problem into its factorable form and relaxing each of the simple functional forms present in the model. For example, see Algorithm Relax f(x) in Tawarmalani and Sahinidis~\cite{ts:comp:04}.

\begin{figure*}[htbp]
\centering
\includegraphics[scale = 0.4]{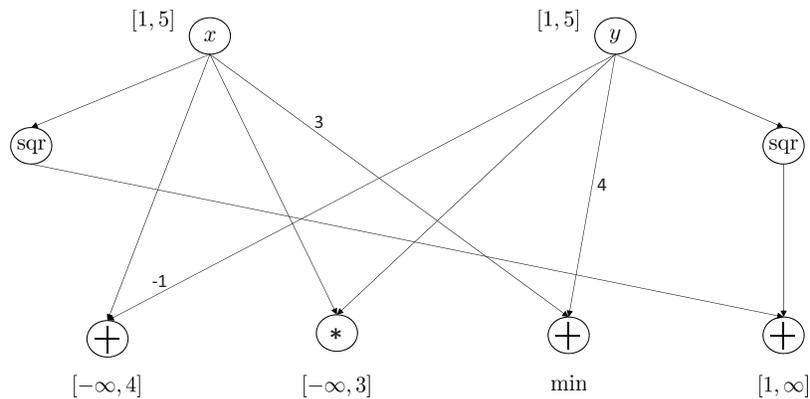}
\caption{Directed Acyclic Graph representation for Problem~\ref{ex}}
\label{fig:dag}
\end{figure*}

Factorable reformulations can be conveniently represented through a directed acyclic graph~\cite{sp96,sn05}. In this graph representation, variables ($x,y,z$) and constants are leaf nodes, vertices are elementary operations ($+, -, *, /$, $\log$, $\exp$, etc.) and the functions to be represented are the root nodes. Variable and constraint bounds are represented through suitable intervals for the root and leaf nodes. Common subexpressions are combined to reduce the size of the graph as doing so is known to tighten the resulting relaxations~\cite[Theorem 2]{ts:comp:04}. Different mathematical formulations can be generated from the same DAG depending on the needs of the solver. Expressions are evaluated by propagating values from the leaves to the root node through the edges in a forward mode. Backward propagation is utilized for the computation of slopes and derivatives. Slopes can also be utilized to construct linear relaxations for the problem. Figure~\ref{fig:dag} represents the DAG for Problem~\ref{ex}.

\section{Interval arithmetic}
\label{sec:int:ar}
Interval arithmetic is a system of arithmetic based on intervals of real numbers~\cite{mkc09}. An interval variable is defined using a variable's lower and upper bounds; the variable itself is restricted to lie between the bounds. Consider the interval variables $\vec{x} = [\vec{x}^l, \vec{x}^u]$ and $\vec{y} = [\vec{y}^l, \vec{y}^u] $. Addition on two intervals can be defined as:
\begin{equation}
\vec{x} + \vec{y} = [\vec{x}^l + \vec{y}^l, \vec{x}^u + \vec{y}^u]
\label{eq:introadd}
\end{equation}
The addition operator defined by equation~\ref{eq:introadd} has the following property:
\begin{equation}
\vec{x} + \vec{y} = \{x + y \hspace{0.2cm} | \hspace{0.2cm} x \in \vec{x} \text{  and  } y \in \vec{y}\}
\end{equation}
Extensions to the definitions of other classic operators can be similarly defined:
\begin{align}
\vec{x} - \vec{y} & = [\vec{x}^l - \vec{y}^u, \vec{x}^u - \vec{y}^l]\\
\vec{x} \times \vec{y} & = [\min (\vec{x}^l \times \vec{y}^l, \vec{x}^l \times \vec{y}^u, \vec{x}^u \times \vec{y}^l, \vec{x}^u \times \vec{y}^u)]\\
\frac{1}{\vec{x}}& = [1/\vec{x}^u, 1/\vec{x}^l], \quad 0 \notin [\vec{x}^l, \vec{x}^u]\\
\frac{\vec{x}}{\vec{y}} &= \vec{x} \times 1/\vec{y}, \quad 0 \notin [\vec{y}^l, \vec{y}^u]
\end{align}
These operators can be suitably defined to account for infinities within the intervals and for the case when 0 lies in the interval of the denominator in a division operation~\cite{k09}. Natural extensions of factorable functions can be computed by replacing all the elementary operations involved in the computation of a factorable function with their interval counterparts. A natural extension provides valid lower and upper bounds for the value of a function.

Floating point arithmetic is susceptible to rounding errors, which can lead to solutions being lost even for simple problems. Neumaier and Shcherbina~\cite{ns04} provide an example of a simple MIP problem that leads to incorrect solutions by major commercial solvers due to roundoff errors. Interval arithmetic methods utilize outward rounding to ensure no points are lost due to roundoff errors. These methods are utilized to design rigorous branch-and-bound-based optimization and root finding methods that ensure no solutions are lost due to roundoff errors~\cite{numerica:97,ar15}. The methods bound function values through interval arithmetic and carry out domain reduction through branching and fathoming tests based on monotonicity, convexity, infeasibility as well as interval Newton type methods which provide conditions for existence and uniqueness of solutions within a box~\cite{h92,n90}.  Interval-based methods have been used for practical applications like solvent blend design~\cite{sag03}. Interval-based solvers include ICOS~\cite{lebbah:09:oms}, Globsol~\cite{kearfott:09:oms} and Numerica~\cite{numerica:97}. Interval-based solvers are often slower than their nonrigorous counterparts. More recently, aspects of safe computations are being introduced with different parts of nonrigorous optimization algorithms.  For example, Neumaier and Shcherbina~\cite{ns04} describe methods to guarantee safety in the solution of MIPs through suitable preprocessing of the LP relaxations and postprocessing of their solutions. Their methods lead to valid results even when the LP solver itself does not use rigorous methods for obtaining a solution. Borradaile and van Hentenryck~\cite{bv05} provide ways of constructing numerically safe linear underestimators for univariate and multivariate functions. Numerically safe methods for computation have also been developed by the CP community, and are referenced in the remainder of the paper including in Sections~\ref{sec:fbbt} and \ref{sec:subopt}.

\section{Presolving optimization models}
\label{sec:presolve}
The idea of analyzing and converting an optimization model into a form more amenable to fast solution is old. Starting with the work of Brearley et al.~\cite{bmw75}, a number of techniques have been used for analyzing models in the operations research community. We use the term presolve to denote all the techniques used for simplification of optimization models. These techniques have been developed extensively for linear programming models~\cite{tw83,tw86,iv96,g97:lp} and for mixed-integer linear programming models~\cite{vw87,hk91,nss94,sav94,m10}. Bixby and Rothberg~\cite{br07} observe that turning off root node presolve at the start of a branch-and-bound search degrades performance of CPLEX 8.0 by a factor of 10.8, while turning off presolve at every node other than the root node degrades the performance by a factor of 1.3 on certain MIP models, demonstrating the importance of presolve. See Achterberg and Wunderling~\cite{aw13} for an extensive computational analysis of the impact of various components of presolve algorithms for MIPs. Some of the ideas for simplification of models include:
\begin{itemize}
\item Elimination of redundant constraints
\item Identification and elimination of dominated constraints (dominated constraints are constraints with a feasible region that is a superset of the feasible region of other constraints in the model)
\item Elimination of redundant variables
\item Assimilating singleton rows into bounds on variables
\item Tightening bounds on dual variables
\item Fixing variables at their bounds
\item Increasing sparsity in the model
\item Rearrangement of variables and constraints to induce structure
\end{itemize}

Similar preprocessing operations can be derived for nonlinear problems~\cite{abg92,l92,fm01,ms03,gt04}. Some of the general guidelines include:
\begin{itemize}
\item Avoid potentially undefined functions
\item Reduce nonlinearity in the model
\item Improve scaling of model
\item Increase convexity in the model through reformulations
\end{itemize}

Amarger et al.~\cite{abg92} provide a software implementation REFORM to carry out many of these reformulations. Presolve techniques are usually implemented at the solver level by developers of various software for optimization. The modelling systems AIMMS~\cite{aimms} and AMPL~\cite{ampl} also provide dedicated presolve systems that are applied to all optimization models~\cite{fg94,h11}. The success of presolve in mathematical programming has led to efforts for its extension to general constraint programming such as automated reformulation of CP models~\cite{lt15,o10:AI}. These methods can lead to considerable simplifications in the model, and can also lead to a reduction in the memory required to solve the problem. Another advantage of these presolve methods is that they are often able to detect infeasibility in optimization models. If a presolved model is infeasible, then the original model is also infeasible. Presolve methods can also be used to detect and correct the causes of infeasibilities for infeasible optimization models. Chinneck~\cite{c08} provides examples of simple cases where infeasibilities can be correctly diagnosed by analyzing the sequence of reductions obtained with presolve. More recently, Puranik and Sahinidis~\cite{ps16} provide an automated infeasibility diagnosis methodology through the isolation of irreducible inconsistent sets (IISs). An IIS is defined as an infeasible set of constraints that has every subset feasible. Isolating an IIS can help accelerate the process of model correction by allowing the model expert to focus onto a smaller problem area within the model. The authors of~\cite{ps16} propose a deletion presolve procedure that exploits feasibility-based bounds tightening techniques in order to accelerate the isolation of an IIS.

While simplified models are often easier to solve than their original counterparts, once an optimal solution has been obtained, the modelling system must return a solution that can be interpreted by the user with respect to the original model form that was specified. Restoration procedures to obtain primal and dual solutions to the original problem are described by Andersen and Andersen~\cite{aa95old} and Fourer and Gay~\cite{fg94}. Such procedures are not discussed here.  In the subsequent sections, we review domain reduction techniques that are a major component of all presolve algorithms.

\section{Reduction of infeasible domains}
\label{sec:fbbt}

The methods described in this section eliminate regions from the search space where no feasible points of Problem~\ref{eq:prob} can exist. Global optimization algorithms maintain continuous and discrete variable domains through upper and lower bounds. Domain reduction is achieved by making these bounds tighter. Therefore, domain reduction is also commonly referred to as bounds tightening.

The tightest possible bounds based on feasibility of the constraints of Problem~\ref{eq:prob} can be obtained by solving the following problems for each of the $n$ variables ($k = 1,\ldots,n$):
\begin{equation} \label{eq:obbt}
\left. \begin{array}{lll}
\min \;\;& \pm x_{k}  \\
\st &  \vec{g(\vec{x})} \leq 0  \\
 &  \vec{x}^{l}\leq \vec{x} \leq \vec{x}^{u}  \\
 & \vec{x} \in \mathbb{R}^{n-m}\times \mathbb{Z}^m
 \end{array} \right\}
\end{equation}
$\pm x_{k}$ in problem~\ref{eq:obbt} denotes two optimization problems, where $x_k$ and $-x_k$ are individually minimized. Solution of these optimization problems returns the tightest possible bounds on the feasible region. If these bounds are tighter than the user-specified bounds in the model, we can achieve domain reduction by using them. However, since the constraints of Problem~\ref{eq:obbt} are potentially nonconvex and due to the presence of integer variables, these are expensive global optimization problems, which might be as hard to solve as Problem~\ref{eq:prob}. A computationally inexpensive way as compared to problem~\ref{eq:obbt} to obtain valid bounds for each of the variables is by solving the following problems for each of the $n$ variables ($k = 1,\ldots,n$):
\begin{equation} \label{eq:obbt-conv}
\left. \begin{array}{lll}
\min \;\;& \pm x_{k}  \\
\st &  \vec{g}_{\rm conv}(\vec{x}) \leq 0  \\
 &  \vec{x}^{l}\leq \vec{x} \leq \vec{x}^{u}  \\
 & \vec{x} \in \mathbb{R}^{n}
 \end{array} \right\}
\end{equation}
where $\vec{g}_{\rm conv}(\vec{x}) \leq 0$ refers to a convex relaxation of the constraints. The convex relaxation can be nonlinear. The integrality restrictions on the variables are relaxed as well. While bounds obtained from Problem~\ref{eq:obbt-conv} in general are weaker than the bounds obtained from Problem~\ref{eq:obbt}, they require the solution of convex optimization problems and are therefore obtained more efficiently. Convex relaxations are utilized in branch-and-bound methods for obtaining valid lower bounds to the optimum, and are thus already available for bounds tightening. To exploit the efficiency and robustness of LP solvers, these convex relaxations are linearized through outer approximation to obtain linear relaxations~\cite{ts:convpoly:05}. Linear relaxations may provide weaker bounds than nonlinear ones, but can be solved very efficiently. This linearization can also be utilized for obtaining tighter bounds on variables through the solution of the following problems for each of the $n$ variables ($k = 1,\ldots,n$):
\begin{equation} \label{eq:obbt-lin}
\left. \begin{array}{lll}
\min \;\;& \pm x_{k}  \\
\st &  \vec{g}_{\rm lin}(\vec{x}) \leq 0  \\
 &  \vec{x}^{l}\leq \vec{x} \leq \vec{x}^{u}  \\
 & \vec{x} \in \mathbb{R}^{n}
 \end{array} \right\}
\end{equation}
where $\vec{g}_{\rm lin}(\vec{x}) \leq 0$ refers to a linearized outer approximation of the feasible region. The bounds obtained from Problem~\ref{eq:obbt-lin} are in general weaker than the bounds obtained from Problem~\ref{eq:obbt-conv}, but require the solution of linear instead of nonlinear programming problems and are therefore solved more efficiently.

Feasibility-based arguments for reduction can also be utilized for tightening constraints. Consider a linear set of constraints $\vec{Ax} \leq \vec{b}$. Tight bounds on constraint $i$ can be obtained with the solution of following optimization problems~\cite{sav94}:
\begin{equation} \label{eq:obbt-constraints}
\left. \begin{array}{lll}
\min \;\;& \pm \vec{a}_i^T\vec{x} \\
\st &  \vec{a}_j^T\vec{x} \leq b_j \qquad j = 1,\ldots,i-1, i+1,\ldots,n   \\
 &  \vec{x}^{l}\leq \vec{x} \leq \vec{x}^{u}  \\
 & \vec{x} \in \mathbb{R}^{n}
 \end{array} \right\}
\end{equation}
Problem~\ref{eq:obbt-constraints} can help identify redundancy if the maximum value of $\vec{a}_i^T\vec{x}$ is strictly less than $b_i$. Conversely, if the minimum value of $\vec{a}_i^T\vec{x}$ is strictly greater than $b_i$, this also identifies infeasibility.

In general, full solution of LP or NLP problems for bounds tightening can be expensive. To balance the computational effort involved with the reduction in bounds obtained, these techniques are usually carried out only at the root node and/or for a subset of variables. They are utilized only sparingly through the rest of the search or not at all. In some cases, however, solving optimization problems for tightening methods throughout the search has been shown to provide significant computational benefits~\cite{cg14}.

Note that reduction methods exploiting Problems~\ref{eq:obbt}, \ref{eq:obbt-conv} or \ref{eq:obbt-lin} are often referred to in the literature as optimality-based bounds tightening. While these methods utilize the solution of optimization problems, they only carry out reduction of infeasible regions from the search space. For this reason, we prefer to classify them under feasibility-based reduction techniques.

\subsection{Bounds propagation techniques}
\label{subsec:fbbt}

Propagation-based bounds tightening techniques will be simply referred to as propagation in the remainder of the paper.  These techniques find their roots in several works in the literatures of mathematical programming~\cite{bmw75}, constraint logic programming~\cite{c87}, interval arithmetic~\cite{cl84}, and AI~\cite{d87}. These methods are often referred to as bounds propagation techniques in CP and are a specialization of constraint propagation. Davis~\cite{d87} refers to a constraint network with nodes (variables) which can take possible labels (domains) and are connected by constraints. He further refers to six different categories of constraint propagation based on the type of information which is updated:
\begin{itemize}
\item Constraint inference: New constraints are inferred and added.
\item Label inference: Constraints are utilized to restrict the sets of possible values for nodes.
\item Value inference: Nodes are partially initialized and constraints are utilized to complete assignments for all nodes.
\item Expression inference: Nodes are labelled with values expressed over other nodes.
\item Relaxation: Nodes are assigned exact values which may violate certain constraints.
\item Relaxation labelling: Nodes are assigned labels using probabilities. Updates in the network involve update of the probabilities.
\end{itemize}
Propagation is equivalent to label inference in the AI community. Davis utilizes the Waltz algorithm~\cite{w75} to describe bounds propagation in a constraint network. In the CP community, these methods are often utilized for solving discrete constraint satisfaction problems (CSP)~\cite{h07} with backtracking-based search methods~\cite{v06a}. However, they are also utilized for continuous CSPs~\cite{bo97,d87}.

Hager~\cite{h93} discusses a reduce operator to eliminate regions outside the solution set for solving systems of constraints. These methods have also been developed extensively in the mathematical programming community, first for LPs, and later for nonlinear programming problems~\cite{thakur90,hansen+:91,hm93,lamar93,m04}. Mclinden and Mangasarian~\cite{mm85} demonstrate inference of bounds for simple monotonic complementarity problems and convex problems. Lodwick~\cite{l89} analyzed the relationship between the constraint propagation methods from AI and bound tightening methods from mathematical programming. These methods typically operate by systematically analyzing one constraint at a time to infer valid bounds on variables.

Propagation techniques are computationally inexpensive. For example, consider a set of linear constraints:
\begin{align*}
\notag \sum_{j=1}^{n}a_{ij}x_j\leq & b_i  \qquad \qquad i = 1,...,k\\
\vec{x}^l \leq \vec{x} \leq & \vec{x}^u
\end{align*}
The following inequalities are implied by every linear constraint:
\begin{equation}
 x_h \leq \frac{1}{a_{ih}}(b_i - \sum_{j \neq h} \min { \left(a_{ij}x_j^U,a_{ij}x_j^L \right)}), \qquad a_{ih} > 0
 \label{ineq:1}
\end{equation}
\begin{equation}
 x_h \geq \frac{1}{a_{ih}}(b_i - \sum_{j \neq h} \min { \left(a_{ij}x_j^U,a_{ij}x_j^L\right) }), \qquad a_{ih} < 0
 \label{ineq2}
\end{equation}
If these inequalities imply tighter bounds on $x_h$ than the ones specified by the model, the bounds can be updated. For example, consider the set of inequalities:
\begin{align}
x_1 + x_2 & \geq 4  \label{eq:ex1}\\
x_2 + x_3 & \leq 1  \label{eq:ex2}\\
x_1 \in& [-2, 4] \notag \\
x_2 \in &[0, 4] \notag \\
x_3 \in& [-1, 1]\notag
\end{align}

From inequality~\ref{eq:ex1}, we can infer $x_1 \geq 4 - \max(0, 4)= 0$. Thus, the lower bound of $x_1$ is updated to 0. From inequality~\ref{eq:ex2}, we can infer that $x_2 \leq 1 - \min(-1, 1) = 2$. Thus, the upper bound of $x_2$ is updated to 2. By analyzing inequality~\ref{eq:ex1} again, we can update the lower bound of $x_1$ to 2. The domains of the variables after these bounds tightening steps are $x_1 \in [2, 4], x_2 \in [0, 2] \text{ and } x_3 \in [-1, 1]$. Harvey and Schimpf~\cite{hs02} describe how bounds can be iteratively tightened in sublinear time for long linear constraints with many variables.

Inequalities~\ref{ineq:1} and~\ref{ineq2} indicate that only one of the bounds for a variable can be updated from an inequality constraint based on the sign of its coefficient. Achterberg~\cite{a09:thesis} formalizes this through the concept of variable locks. Thus, inequality~\ref{eq:ex1} down locks variables $x_1$ and $x_2$, since the lower bounds for $x_1$ and $x_2$ cannot be moved arbitrarily without violating inequality~\ref{eq:ex1}. The concept of variable locks allows for efficient duality fixing of variables. Note that there is no up lock for variable $x_1$ and no down lock for variable $x_3$ based on the inequalities~\ref{ineq:1} and~\ref{ineq2}. Thus, if the coefficient of $x_1$ in the objective function is negative, $x_1$ can be set to its upper bound. Similarly, if the coefficient of $x_3$ in the objective function is positive, $x_3$ can be set to its lower bound. The concept of variables locks was extended for CP to develop presolve procedures for cumulative constraints by Heinz et al.~\cite{hsb13}. Duality fixing is extended by Gamrath et al.~\cite{gkmmw15} to allow for fixing of a singleton column. The authors also define dominance between two variables for a MIP and show how dominance information can be used for fixing variables at their bounds.

\begin{figure*}[htbp]
\centering
\includegraphics[scale = 0.4]{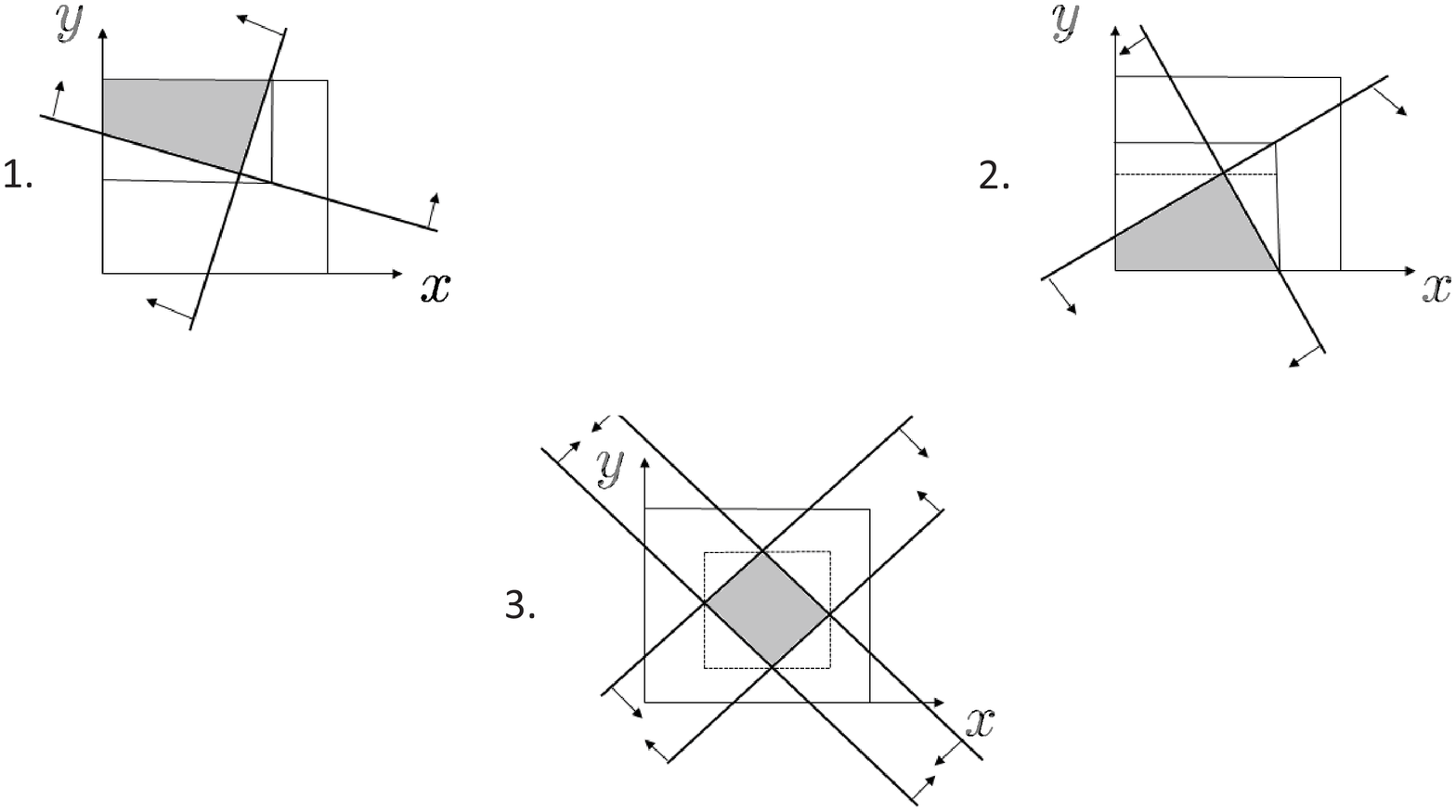}
\caption{Domain reductions through linear propagation}
\label{fig:lbbt}
\end{figure*}

Figure~\ref{fig:lbbt} considers three cases of propagation. In Case 1, bounds for $x$ and $y$ are successfully tightened through iterative application of propagation, with no further tightening possible via Problem~\ref{eq:obbt}. In Case 2, the bound obtained for $x$ by propagation is the tightest, however the bound for $y$ can be tightened further by Problem~\ref{eq:obbt}. In Case 3, bounds for neither $x$ nor $y$ can be tightened by propagation, whereas significant reduction is possible via Problem~\ref{eq:obbt}.

In general, reduction in domain through propagation is not guaranteed. Consider the following example:
\begin{align}
x_1 + x_2 & \geq 0  \notag\\
x_1 + x_2 & \leq 4  \notag\\
-x_1 + x_2 & \geq -2  \notag\\
-x_1 + x_2 & \leq 2  \notag\\
x_1 \in& [-3, 5] \notag \\
x_2 \in &[-3, 5] \notag
\end{align}

The tightest possible domain for $x_1$ and $x_2$ based on feasibility arguments is $[-1, 3]$ and can be obtained by solving Problem~\ref{eq:obbt}. However, analyzing constraints one-at-a-time leads to no reduction of bounds for this problem. The reason for this drawback is that propagation only analyzes one constraint at a time, whereas Problem~\ref{eq:obbt} utilizes information from all the constraints in the model simultaneously.

Belotti~\cite{b13} considers bound reduction by generating a convex combination of two linear inequalities and utilizing this combination for bound reduction. The author proposes a univariate optimization problem to determine the optimal value for the convex multipliers. However, the computational complexity of considering all possible combinations of $m$ constraints in $n$ variables is $O(m^2n^3)$. The author proposes a heuristic scheme which, when implemented only on a subset of the nodes of the branch-and-bound tree, shows computational benefits of using this method. For other presolve-based methods for mixed-integer programs analyzing more than one constraint at a time, see Achterberg et al.~\cite{abgrw14}. Domes and Neumaier~\cite{dn16} propose the generation of a new constraint by taking the linear combination or aggregation of all constraints for quadratic problems. This new constraint can uncover new relationships between the variables and can lead to domain reduction. The authors propose the use of the dual multipliers of a local solution in the case of a feasible subproblem and the use of a constraint violation measure in the case of an infeasible subproblem to aggregate constraints. Their method shows computational benefits in reducing the cluster effect (Section~\ref{sec:cluster}). The notion of global constraints in CP~\cite{r11} is related. A global constraint provides a concise representation for a set of individual constraints. Filtering algorithms on a global constraint effectively carry out domain reduction by considering multiple individual constraints simultaneously. Lebbah et al.~\cite{lmr05} present a global constraint for a set of quadratic constraints and describe filtering algorithms. Similar ideas have been extended to polynomial constraints~\cite{lmrdm05}.

Bounds can also be inferred for nonlinear constraints via propagation. Consider a bilinear term of the form $x_i = x_jx_k$. Then,
\begin{equation}
x_i \leq \max \{ x_j^Lx_k^L, x_j^Lx_k^U,x_j^Ux_k^L,x_j^Ux_k^U\} \label{nlineq:1}
\end{equation}
and
\begin{equation}
x_i \geq \min \{ x_j^Lx_k^L, x_j^Lx_k^U,x_j^Ux_k^L,x_j^Ux_k^U\} \label{nlineq:2}
\end{equation}
represent valid bounds for variable $x_i$ and can be used to potentially tighten any user-specified bounds for this variable.

In the context of factorable reformulations, an optimization problem is already decomposed into simple functional forms, which can be readily utilized for domain propagation.  Consider, for instance, the factorable reformulation for Problem~\ref{ex}. From constraint~\ref{eq:non1}, we can infer that $0\leq z_3 \leq 3$. From constraint~\ref{eq:non2}, we can infer that $1\leq y \leq 3$.

The iterative application of simple constraints such as \ref{ineq:1}--\ref{ineq2} and \ref{nlineq:1}--\ref{nlineq:2} is referred to as \textit{poor man's LPs} and \textit{poor man's NLPs}~\cite{ss98,nvs:cocos:03} because this iterative application is an inexpensive way to approximate the solution of linear and nonlinear optimization problems that aim to reduce domains of variables.  Propagation based on these techniques can be applied not only to the original problem constraints but also to any cutting planes and other valid inequalities that might be derived by the branch-and-bound solution algorithm.  Moreover, all these techniques can be implemented efficiently via the DAG representation. If a bound on a variable $x$ has changed, it can be propagated to other variables that depend on $x$ through a simple forward propagation on the DAG based on interval arithmetic. Similarly, the variables on which $x$ depends can be updated through a backward propagation.

\subsection{Convergence of propagation}
\label{fbbt:conv}
Propagation methods can be carried out iteratively as long as there is an improvement in variable bounds. However, these methods can fail to reach a fixed point finitely. Consider as an example~\cite{n04}: $x_1 + x_2 = 0$, $x_1 - qx_2 = 0$ with $q \in (0,1)$. Propagation will converge to $(0, 0)$ at the limit, i.e., as the number of iterations approaches infinity. On the contrary, any linear programming based method will terminate at $(0,0)$ quickly as the only feasible point. A necessary condition for nonconvergence of propagation iterations is the presence of cycles in expression graphs~\cite{f94}. The problem of achieving a fixed point with propagation iterations in the presence of only integer variables has been shown to be NP-complete~\cite{bhv07,bknv11}. The fixed point obtained is independent of the order in which the variables are considered~\cite{cl10}. In practice, the iterations are usually terminated when the improvement in variable domains is insignificant or when an upper limit on the number of iterations is reached.

For linear inequalities in the presence of continuous variables alone, the fixed point can be obtained in polynomial time by the solution of a large linear program~\cite{bcll10}. For nonlinear problems and for problems with integer variables, a linear relaxation can be solved to obtain reduced bounds~\cite{bcll12}.

\subsection{Consistency}
\label{sec:cons}
Constraint programming algorithms for constraint satisfaction problems rely on the formalized notion of consistency to propagate domains along constraint networks and remove values from variable domains that cannot be a part of any solution. Differing levels of consistency exist, including arc consistency and k-consistency, while various algorithms have been proposed to achieve these consistencies~\cite{b06}. These concepts were originally proposed for discrete problems but were also extended to continuous constraint satisfaction problems~\cite{l93,bmp94,sf96,bggp99,cdr99,bsw08}.

The domain reduction algorithms typically employed for global optimization of MINLPs are usually not iterated until they reach a certain level of consistency because attempting to establish consistency can lead to a prohibitively large computational effort. However, domain reduction techniques are not even necessary to prove convergence of branch-and-bound based methods for global optimization. In contrast to CP methods, the availability of relaxation methods for generating bounds allows the mathematical-programming-based branch-and-bound algorithms to converge without establishing any levels of consistency.

\subsection{Techniques for mixed-integer linear programs}
\label{sec:mip}
Many techniques for presolving LPs can also be directly  utilized for MIPs. However, specialized reduction methods can be developed for MIPs. For example, a number of methods have been developed for the analysis of 0-1 binary programs~\cite{gs81,cjp83,sav94}. Probing and shaving techniques are equivalent to the singleton consistency techniques from CP. Singleton consistency techniques~\cite{psw00} proceed by assigning a fixed value to a variable and carrying out propagation. If this assignment leads to infeasibility of the constraint program, the value assigned to the variable cannot be a part of any solution and can be eliminated. Probing techniques, that can be used for binary programs, involve the fixing of a binary variable $x_i$ to say 0. If basic preprocessing and other domain reduction methods are able to prove infeasibility for this subproblem, then the variable $x_i$ can be fixed to 1. If fixing the binary variable $x_i$ to both 0 and 1 lead to infeasibility, the problem can be declared as infeasible. For binary programs, if probing is carried on all binary variables, it leads to singleton consistency. Probing has also been utilized for satisfiability problems~\cite{lm03}. A similar technique for continuous problems is called shaving~\cite{ms96,n04}. The method proceeds by removing a fraction of the domain of a variable $[x_i^l , x_i^u]$ as either $[x_i^l  + \epsilon, x_i^u]$ or $[x_i^l, x_i^u - \epsilon]$ and testing whether the reduced domain leads to provable infeasibility of the model. If infeasibility is indeed proved, the domain of the variable can be reduced to the complement of the box used for testing. The author suggests using $\epsilon = 0.1\times (x_i^u - x_i^l)$ for this method. Shaving is widely used in the sub-field of CP based scheduling~\cite{tl00}. Belotti et al.~\cite{b+:09:oms} call this technique aggressive bounds tightening (ABT) and implement it in the solver COUENNE. Once a fraction of the domain of a variable has been eliminated, propagation is invoked in the hopes of proving infeasibility in the subproblem. However, ABT and shaving techniques in general are expensive and reduction in the domain is not guaranteed. Faria and Bagajewicz~\cite{fb12} propose several variants of this shaving strategy for bilinear terms and utilize it in a branch-and-bound algorithm for water management and pooling problems~\cite{fb11,fb11:2}. Nannicini et al.~\cite{nbllmw11} propose a similar strategy which they refer to as aggressive probing. In contrast to ABT technique, the nonconvex restrictions created by restricting a variable domain in their method are solved to global optimality with branch-and-bound in order to carry out the maximum possible domain reduction.

Implication-based reductions utilize relations between the values of different variables that must be satisfied at an optimal solution. For instance, if fixing two binary variables $x_i$ and $x_j$ to 0 leads to infeasibility, then we have an implication requiring that one of the variables must be nonzero, which can be represented by the inequality $x_i + x_j \geq 1$. Such relations can be efficiently represented through the use of conflict graphs~\cite{ans00}. Conflict graphs are utilized to generate valid inequalities that strengthen the MIP formulation. In general, if fixing a binary variable $x_i$ to 0 implies that variable $x_j$ must take value $v$, then the following inequalities are valid for the problem~\cite{m10}:
\begin{align}
x_j \leq v + (x_j^u - v)x_i  \notag\\
x_j \geq v - (v - x_j^l )x_i \notag
\end{align}
Implications can be derived by carrying out probing by fixing more than one variable at a time and/or through the analysis of problem structure. Implications can also be utilized for identifying and eliminating redundant constraints. Inequalities derived from implications lead to automatic disaggregation for some constraints~\cite{sav94}. Disaggregated constraints, while redundant for the MIP formulation, lead to tighter LP relaxations for the problem. Achterberg et al.~\cite{ass13} show how implications can be derived from conflicts (See Section~\ref{sec:conflict})  and from knapsack covers.

Tighter formulations can also be obtained for a problem through coefficient reduction. Consider the example of a linear constraint on binary variables from~\cite{m01}:
\begin{equation*}
-230x_{10}-200x_{16}-400x_{17} \leq -5
\end{equation*}
The coefficients of this constraint can be reduced to:
\begin{equation*}
-x_{10}-x_{16}-x_{17} \leq -1
\end{equation*}
While the set of binary values satisfying both of the above constraints are the same, the reduced constraint has a tighter LP relaxation. Approaches for coefficient reduction are provided by~\cite{cjp83,sav94}. Andersen and Pochet~\cite{ap10} prove that, if no coefficients for an MIP system can be strengthened, then there does not exist a dominating constraint that can be used to replace an existing constraint in the MIP system to tighten its relaxation. They also describe an optimization formulation and an algorithmic solution to the problem of strengthening a coefficient in a constraint as much as possible.

\subsection{Conflict analysis}
\label{sec:conflict}
Branch-and-bound based algorithms often encounter infeasibility in subproblems during the search for global optimum. Solvers for the satisfiability problem (SAT) also utilize a backtracking-based branching scheme for their solution~\cite{mmzzm01}. The SAT problem consists of binary variables constrained by a set of logical conditions. The SAT problem has a solution if there exists an instantiation of the binary variables satisfying all the logical conditions. SAT-based solvers learn and add conflict clauses from infeasible subproblems~\cite{ms99}. These clauses are created by identifying the corresponding instantiations of a subset of variables leading to infeasibility and prohibiting them. This often leads to a reduction in the search tree. The idea of conflict analysis is similar to the idea of no-good learning from the CP community~\cite{ss77,lstv07,k08,ksv15}. No-good is a generalization of a conflict clause from SAT to CP.

The ideas of conflict analysis have been extended for MIP~\cite{a07,ss06}. However, generation of conflict clauses is more complicated for MIP due to the presence of both continuous and general integer variables. Infeasibility in SAT problems and CP problems is detected through a chain of logical deductions caused by fixing some variables. However, in MIP, infeasibility can be identified through either such deductions or an infeasible LP relaxation. Achterberg~\cite{a07} proposes a generalized conflict graph (termed implication graph in~\cite{ss06}) for representation of bound propagations that can be used to identify cause of infeasibility. Note that the conflict graph and the implication graph are defined differently than in Section~\ref{sec:mip}. In the case of an infeasible LP relaxation, Achterberg proposes identifying a minimum bound-cardinality IIS. Representation of conflict causes for MIP requires use of disjunctive constraints which must be reformulated with additional binary variables and inequalities. Limited computational analysis demonstrates a reduction in the number of nodes and time with conflict analysis.

\section{Reduction of suboptimal domains}
\label{sec:subopt}

In contrast to the methods of the previous section, the methods described here can lead to the elimination of feasible points from the domain under the condition that at least one globally optimal solution remains within the search space.

\begin{figure*}[htbp]
\centering
\includegraphics[scale = 0.5]{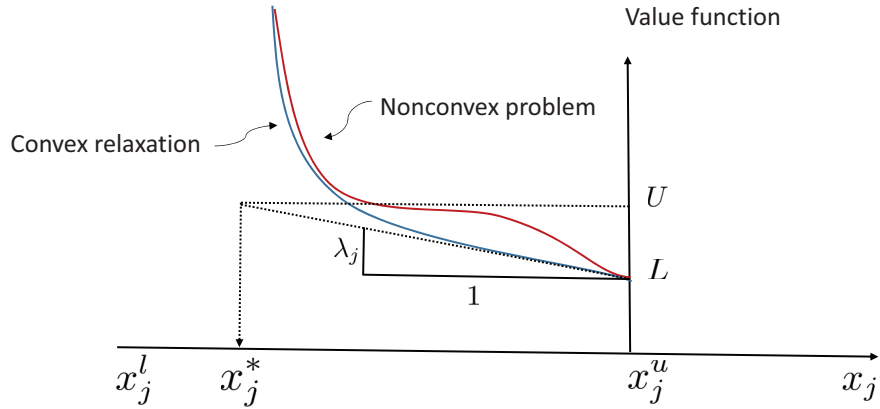}
\caption{Multipliers-based reduction geometrically}
\label{fig:dbr}
\end{figure*}

Consider the following convex relaxation of Problem~\ref{eq:prob}:
\begin{equation} \label{eq:probing-conv}
\left. \begin{array}{lll}
\min \;\;& f_{\text{conv}}(\vec{x})  \\
\st &  \vec{g}_{\rm conv}(\vec{x}) \leq 0  \\
 &  \vec{x}^{l}\leq \vec{x} \leq \vec{x}^{u}  \\
 & \vec{x} \in \mathbb{R}^{n}
 \end{array} \right\}
\end{equation}
As discussed before, convex relaxations are often constructed and linearized in a global branch-and-bound search for obtaining valid lower bounds.  Assume that the optimal objective for Problem~\ref{eq:probing-conv} has value $L$ and, at the optimal solution, a bound $x_j \leq x_j^u$ is active with a Lagrange multiplier $\lambda_j > 0$. Let $U$ be a known valid upper bound for the optimal objective function value of Problem~\ref{eq:prob}. Then, the following constraint does not exclude any optimal solutions better than $U$ (Theorem 2 in~\cite{rs95}):
\begin{equation}
x_j^l \geq x_j^u - \frac{U-L}{\lambda_j}
\label{eq:prob1}
\end{equation}
A geometric interpretation of this cut can be observed in Figure~\ref{fig:dbr}, where $x_j^*$ denotes the right-hand-side of equation~\ref{eq:prob1}. The constraint excludes values of $x_j$ for which the convex relaxation is guaranteed to have its value function to be greater than or equal to $U$. Consequently, the nonconvex problem also has its value function guaranteed to be greater than or equal to $U$ in this domain. A corresponding cut can also be derived if, at the optimal solution $L$ of the convex relaxation, a variable is at its lower bound, i.e., $x_j = x_j^l$, with the corresponding Lagrange multiplier $\lambda^j > 0$. The upper bound can then be potentially tightened without losing optimal solutions by the following cut:
\begin{equation}
x_j^u \leq x_j^l + \frac{U-L}{\lambda_j}
\label{eq:prob2}
\end{equation}

For variables that are not at their bounds at the relaxation solution, probing tests can be carried out by temporarily fixing variables at their bounds and solving the restricted problem. For example, Tests 3 and 4 in~\cite{rs95} work as follows.  Set $x_j = x_j^u$ and solve Problem~\ref{eq:probing-conv}. If the corresponding multiplier  $\lambda^j $ is positive, then the following constraint is valid:
\begin{equation}
x_j^l \geq x_j^u - \frac{U-L}{\lambda_j}
\label{eq:prob3}
\end{equation}
A similar probing test can be developed by fixing a variable at its lower bound~\cite{rs95,rs96,ss98}. Reduction by~\ref{eq:prob1} and~\ref{eq:prob2} only requires the solution of the relaxation problem, and thus can be implemented at every node of the branch-and-bound tree without much computational overhead. On the other hand, probing with fixing variables at their bounds requires the solution of $2n$ convex problems. Probing is usually carried out only at the root node and then only at some nodes of the tree or only for a subset of the variables. The probing tests are a generalization of probing for the integer programming case. However, they differ from the integer programming case in the sense that probing leads to variable fixing in the case of binary variables, whereas in the continuous case it leads to reduction in the variable domains.

Similarly, if at the optimal solution of the relaxation Problem~\ref{eq:probing-conv}, a constraint $g^j_{\rm conv}(\vec{x}) \leq 0$ is active with the corresponding multiplier $\mu_i \geq 0$, then the following constraint does not violate any points with objective values better than $U$~\cite{rs96}:
\begin{equation}
\label{eq:prob:const}
g^j_{\rm conv}(\vec{x}) \geq  - \frac{U-L}{\mu_j}
\end{equation}
These inequalities can be appended to the original formulation.  However, this process can lead to the accumulation of a large number of constraints. Alternatively, these are often utilized for propagation-based tests locally at the current node and then discarded. Lebbah et al.~\cite{lmr07} provide a safe way of implementing duality-based reduction techniques in the context of floating point arithmetic. It must be noted that suboptimal Lagrangian multipliers often provide greater reduction through constraints \ref{eq:prob1}, \ref{eq:prob2}, \ref{eq:prob3} and \ref{eq:prob:const}. A more detailed discussion on the use of other than optimal multipliers is provided in Tawarmalani and Sahinidis~\cite{ts:comp:04} and Sellmann~\cite{s04}.

If a valid upper bound \emph{U} is available, it can also be exploited by reformulating Problem~\ref{eq:obbt-conv} as follows:
\begin{equation} \label{eq:obbt-cont-conv}
\left. \begin{array}{lll}
\min \;\;& \pm x_{k}  \\
\st &  \vec{g(\vec{x})} \leq 0  \\
&  f_{\text{conv}}(\vec{x}) \leq U  \\
 &  \vec{x}^{l}\leq \vec{x} \leq \vec{x}^{u}  \\
 & \vec{x} \in \mathbb{R}^{n}
 \end{array} \right\}
\end{equation}
Problem~\ref{eq:obbt-cont-conv} differs from Problem~\ref{eq:obbt-conv} by the addition of a single constraint. In this constraint, $f_{\text{conv}}(\vec{x})$ refers to the convex relaxation of the objective function. It can be replaced by a suitable linearization  $f_{\text{lin}}(\vec{x})$ for use in formulation~\ref{eq:obbt-cont-lin}.
\begin{equation} \label{eq:obbt-cont-lin}
\left. \begin{array}{lll}
\min \;\;& \pm x_{k}  \\
\st &  \vec{g}_{\rm lin}(\vec{x}) \leq 0  \\
&  f_{\text{lin}}(\vec{x}) \leq U  \\
 &  \vec{x}^{l}\leq \vec{x} \leq \vec{x}^{u}  \\
 & \vec{x} \in \mathbb{R}^{n}
 \end{array} \right\}
\end{equation}
Zamora and Grossmann~\cite{zg99} refer to Problem~\ref{eq:obbt-cont-conv} as the contraction subproblem. They define a contraction operation that uses the solution of Problem~\ref{eq:obbt-cont-conv} along with multipliers-based reduction steps described in this section to carry out domain reduction. Their algorithm is therefore referred to as branch-and-contract. Optimality-based arguments can also be used for filtering in CP, see for example~\cite{flm99,to02}.

Note that problem~\ref{eq:obbt-cont-conv} can be solved iteratively to obtain further tightening. However, convergence to a fixed point can be slow. Caprara and Locatelli~\cite{cl10} propose a different approach to carry out bounds tightening called nonlinearities removal domain reduction (NRDR). NRDR relies on the solution of parametric univariate optimization problems to find tight bounds. The method reduces nonlinearities due to a single variable at every iteration. The authors further demonstrate that, under certain assumptions, their domain reduction method is equivalent to carrying out optimization-based bounds tightening iteratively until it reaches a fixed point. Caprara et al.~\cite{cmm16} show that the NRDR method leads to bounds consistency for a special case of linear multiplicative programming problems with only two variables in the objective function.

\subsection{A unified theory for feasibility- and optimality-based tightening}

Tawarmalani and Sahinidis~\cite{ts:comp:04} provide a unified theory of feasibility- and optimality-based bounds tightening techniques.  This theory evolves around Lagrangian subproblems created by the dualization of constraints.  Consider the problem:
\begin{equation} \label{eq:nlpprob}
\left. \begin{array}{lll}
\min  \;\;& f(\vec{x}) \\
\st & \vec{g(\vec{x})} \leq 0 \\
 & \vec{x}_{l}\leq \vec{x} \leq \vec{x}_{u} \\
 & \vec{x} \in \mathbb{R}^{n}
\end{array} \right\}
\end{equation}
Define the Lagrangian subproblem as follows:
\begin{equation*}
\inf_{\vec{x}_{l}\leq \vec{x} \leq \vec{x}_{u} }\{-y_0 f(\vec{x}) - \vec{y} \vec{g(x)}\}
\end{equation*}
The dual variables $y_0$ and $y$ are nonpositive. Suppose an upper bound $U$ is available for the optimal solution of Problem \ref{eq:nlpprob}. A domain reduction master problem can be constructed as follows:
\begin{equation}
\label{eq:red-master}
\left. \begin{array}{lll}
   \inf \;\; & h(\mu_0, \vec{\mu}) \\
 \st &  f(\vec{x}) \leq \mu_0 \leq U \\
 &  \vec{g}(\vec{x}) \leq \vec{\mu} \leq b\\
&  \vec{x}_{l}\leq \vec{x} \leq \vec{x}_{u}
\end{array} \right\}
\end{equation}
Using the linear objective function $h(\mu_0, \vec{\mu}) = a_0\mu_0 + \vec{a}^T\vec{\mu}$, Problem \ref{eq:red-master} can be equivalently stated as:
\begin{equation} \label{eq:red-master-eqv}
\left. \begin{array}{lll}
\inf  \;\; &a_0\mu_0 + \vec{a}^T\vec{\mu}  \\
 \st & -y_0(f(\vec{x}) - \mu_0 )\\
     & \quad -\vec{y}^T(\vec{g}(\vec{x})-\mu) \leq 0 \quad \forall(y_0,\vec{y}) \leq 0\\
& (\mu_0,\vec{\mu}) \leq (U,0) \\
&  \vec{x}_{l}\leq \vec{x} \leq \vec{x}_{u}
\end{array} \right\}
\end{equation}
Solving Problem \ref{eq:red-master-eqv} can be as hard as solving Problem \ref{eq:nlpprob}. Instead, the lower bound for Problem \ref{eq:red-master-eqv} can be obtained from the solution of the following relaxed master problem:
\begin{equation} \label{eq:red}
\left. \begin{array}{lll}
\inf  \;\;& a_0\mu_0 + \vec{a}^T\vec{\mu} \\
\st & y_0\mu_0 + \vec{y}^T\vec{u} \\
    & \quad + \inf_{\vec{x}_{l}\leq \vec{x} \leq \vec{x}_{u}}\{-y_0f(\vec{x}) - \vec{y}^T\vec{g(x)}\}\\
    & \quad \leq 0  \quad \forall (y_0,\vec{y}) \leq 0\\
& (\mu_0,\vec{\mu}) \leq (U,0)
\end{array} \right\}
\end{equation}
Many domain reduction operations can be obtained by suitable choice of the coefficients $(a_0, \vec{a})$ in the objective function for Problem \ref{eq:red}. For example, equation \ref{eq:prob:const} can be derived by setting $(\vec{a},a_0)$ to $\vec{e}_j$ in Problem \ref{eq:red}, where $\vec{e}_j$ is the $j^{\rm th}$ column of the identity matrix.

Similarly, propagation for linear constraints described by equations \ref{ineq:1} and \ref{ineq2} can derived by application of duality theory to the relaxed problem formed by relaxing all but the $i^{\rm th}$ linear constraint under consideration:
\begin{align*}
\min  \quad & x_h \\
\st \quad & \vec{a}_i^T \vec{x} \leq b_i  \\
& \vec{x} \leq \vec{x}^{u}  \\
& \vec{x} \geq \vec{x}^l
\end{align*}
Assuming $a_{ih} < 0$ and $\vec{x}_h$ is not at its lower bound, the optimal dual solution for the linear program can be obtained as:
\begin{align*}
\mu = &1/a_{ih}  \\
\vec{\lambda}_j = &-\max\{a_{ij}/a_{ih}, 0\} \text{ for all } j \neq h  \\
\vec{\sigma}_j = &\min\{a_{ij}/a_{ih}, 0\} \text{ for all } j \neq h  \\
\vec{\lambda}_h = & \vec{\sigma}_h = 0
\end{align*}
Here, $\mu$ is the dual multiplier corresponding to $\vec{a}_i^T \vec{x} \leq b_i$, $\vec{\lambda}_j$ is the dual multiplier corresponding to $\vec{x}_j \leq \vec{x}_j^u $, and $\vec{\sigma}_j$ is the dual multiplier corresponding to $\vec{x}_j \geq \vec{x}_j^l $. The domain reduction master problem can be constructed as:
\begin{equation} \label{eq:dual:prop}
\left. \begin{array}{lll}
\min \;\;& x_h \\
\st & - x_h + \mu u + \vec{\lambda}^T \vec{v} - \vec{\sigma}^T \vec{w} \leq 0 \\
& u \leq b_i \\
& \vec{v} \leq \vec{x}^u \\
& \vec{w} \leq \vec{x}^l
\end{array} \right\}
\end{equation}
Equation \ref{ineq2} follows from Problem \ref{eq:dual:prop}. Equation \ref{ineq:1} can be similarly derived.

Tawarmalani and Sahinidis~\cite{ts:comp:04} also present a duality-based reduction scheme that utilizes dual feasible solutions and a learning reduction heuristic. In branch-and-bound algorithms, branching is typically carried out by various heuristics for the selection of the branching variable and the branching point. If one of the nodes created due to a branching decision is proven to be inferior, the learning reduction heuristic attempts to expand on the region defined by this node that is proven to be inferior by the construction of dual solutions for the other node. The authors remark that all dual feasible solutions can be utilized to carry out reductions. This idea is instantiated in the Lagrangian variable bounds propagation by Gleixner and coworkers~\cite{gbmw16,gw13}. To avoid solving $2n$ optimization problems at every node with Problem~\ref{eq:obbt-cont-lin}, valid Lagrangian variable inequalities can be generated by aggregating the linear relaxation constraints with the dual solution of Problem~\ref{eq:obbt-cont-lin}. While redundant for the linear relaxation, these inequalities approximate the effect of solving Problem~\ref{eq:obbt-cont-lin} locally at every node and can be used to infer stronger bounds on variables when variable bounds are updated through branching or otherwise if a better upper bound is obtained. Gleixner et al.~\cite{gbmw16} provides heuristics for ordering the generated Lagrangian variable inequalities to achieve maximum tightening. An aggressive filtering strategy is proposed that involves the solution of LPs to determine variables for which bounds cannot be tightened. The optimization steps for these variables in Problem~\ref{eq:obbt-cont-lin} can therefore be skipped, leading to computational savings. The authors also provide heuristics for ordering the LP solves in Problem~\ref{eq:obbt-cont-lin} to allow for more efficient warm starting and reduce the number of simplex iterations.

\subsection{Pruning}
If the lower bound obtained at a node of the branch-and-bound search is worse than the best known upper bound $U$, the current node can be fathomed. We are guaranteed that the globally optimal solution cannot lie at this node, since all feasible solutions at this node have an objective value greater than $U$. This process is also referred to as pruning and is a special case of a general concept called dominance; node $n_1$ dominates node $n_2$ if, for every feasible solution $s$ in $n_2$, there is a complete solution in $n_1$ that is as good or better than $s$. The idea of dominance is old, first introduced by Kohler and Steiglitz~\cite{ks74} and developed in more detail by Ibaraki~\cite{i77}. Dominance relations can be utilized to speed up the search. For MIP problems, Fischetti and Toth~\cite{ft88} propose solution of an auxiliary MIP involving only fixed variables at a node to determine whether the current node is dominated by another node which may or may not have been explored yet. However, the overhead of solving the auxiliary MIP is fairly large. Fischetti and Salvagnin~\cite{fs10} propose improvements to this scheme and demonstrate computational benefits for network loading problems arising in telecommunication. Sewell et al.~\cite{ssmjk12} propose a memory-based dominance rule for a scheduling application. Memory-based dominance rules require storage of the entire search tree, and their performance is dependent on the memory available. However, since information from the entire tree is available, considerably stronger pruning rules can be determined. Memory-based search algorithms are related to the heuristic search algorithms from AI~\cite{es11,s12}.

Problems involving integer variables often have a high degree of symmetry. Symmetry is detrimental for branch-and-bound algorithms as it can lead to repetitive work for the solver. Most symmetry breaking approaches rely on exploiting information about the specific problem being considered. A more general technique called isomorphic pruning has been proposed by Margot~\cite{m02,m03}. Isomorphic pruning relies on lexicographic tests to determine if a node can be pruned. Orbital branching~\cite{olrs11} is another method that can be utilized to tackle symmetry. The branching method identifies orbits of equivalent variables. Orbital branching proceeds by fixing a variable in the orbit to one at a node, and fixes all the variables in the node to zero in another node. Thus, orbital branching implicitly prunes all the other nodes which involve each of the other variables in the orbit fixed to one.

\subsection{Exploiting optimality conditions}
The Karush-Kuhn-Tucker~\cite{k39,kt51} conditions are necessary for a point to be locally optimal for a nonlinear programming problem under certain constraint qualifications. See Schichl and Neumaier~\cite{sn06} for a derivation and a general discussion of these conditions. The conditions can be used to reduce the search space for a nonconvex NLP, since globally optimal points also satisfy them. Vandenbussche and Nemhauser generate valid inequalities for quadratic programs with box constraints through the analysis of optimality conditions~\cite{vn05a} and also utilize them in a branch-and-cut scheme~\cite{vn05b}. Optimality conditions have been extensively utilized in branch-and-bound algorithms for quadratic programs~\cite{hjrx93,bv08,bv09,bc11,cb12,hmp12}. Optimality conditions can also be utilized for pruning of nodes. For example, for an unconstrained optimization problem, if 0 is not contained within the interval inclusion function for the partial derivatives of the objective function at a node, the corresponding node can be pruned~\cite{ms14}. This test is often referred to as the monotonicity test.

Sahinidis and Tawarmalani~\cite{st:rlxn:05} have added a modelling language construct for BARON which allows for the specification of certain constraints as relaxation-only. Relaxation-only constraints are utilized for the construction of convex relaxations and for domain reduction, but are not utilized in local search for obtaining upper bounds. The authors use first-order optimality conditions explicitly as relaxation-only constraints and observe improved convergence for some univariate optimization problems. Amaran and Sahinidis~\cite{as11} use a similar strategy and show significant computational benefits for parameter estimation problems. They analyze their results and demonstrate that explicit use of optimality conditions aids in domain reduction steps of BARON leading to computational speedups. Puranik and Sahinidis~\cite{ps16:2} propose a strategy for carrying out implicit bounds tightening on optimality conditions for bound-constrained optimization problems. Their method does not require the generation of optimality conditions which can be time consuming and lead to increase in memory requirements. For a large collection of test problems, this strategy leads to computational speedups and reduction in the number of nodes.

\subsection{Cluster effect}
\label{sec:cluster}
Branch-and-bounds methods often undergo repeated branching in the neighbourhood of the global solution before converging. This problem is referred to as the cluster effect and was first studied by Du and Kearfott~\cite{dk94} in the context of interval-based branch-and-bound methods. Subsequent analysis~\cite{n04,ss10,wsb14} also demonstrates the importance of convergence order of the bounding operation (see Definition 1 in~\cite{wsb14}). These results indicate that at least second-order convergence is required to overcome the cluster effect.  Thus, tighter relaxations can indeed help mitigate the cluster effect and their development is the subject of extensive research in the area~\cite{sa90,sa94,ar97,sa99,ts02,mf05,bst:09:oms,ks:11,ks:13,trx13,zs:13:oms,kms:14,bkst:15}. Neumaier and coworkers provide methods to construct exclusion regions for the solution of systems of equations~\cite{sn04} and for the solution of optimization problems~\cite{skm14}. These exclusion regions guarantee that no other solution can lie within the exclusion box around a local minimizer or a solution for systems of equations. These boxes can then be eliminated from the search space.  These boxes are constructed through existence and uniqueness tests based on Krawczyk operator or the Kantorovich Theorem (see Chapter 1,~\cite{k96}). Other methods to construct exclusion regions include back boxing~\cite{v96} and $\epsilon$-inflation~\cite{m95}.

\section{Implementation of domain reduction techniques}
\label{sec:strat}
Domain reduction strategies, if successful, typically lead to a reduction in the number of nodes searched in a branch-and-bound tree. Techniques like propagation are computationally inexpensive and can be applied at every node without much overhead. However, they are not as efficient in carrying out domain reduction as the more computationally intensive strategies like Problem~\ref{eq:obbt} or probing. Thus, there exists a need to balance the effort involved in domain reduction in order to reduce the average effort per node. Multiple heuristic or learning strategies are employed for this purpose. These ideas are important in constraint satisfaction problems where multiple filtering algorithms are available achieving varying degrees of consistency. Stergiou~\cite{s09} experimentally analysed the domain reduction events through filtering techniques. Drawing insights from the clustering of this experimental data, various heuristics are proposed for choosing the propagation algorithm on the fly. Based on insights from Stergiou~\cite{s09} that propagation events often occur in close clusters, Araya et al.~\cite{asc15} propose an adaptive strategy. Thus, if a constraint propagation mechanism succeeds in carrying out domain reduction at a given node, it should be exploited repetitively in other nodes that are geometrically close to the node until the method fails. % the previous sentence is unclear.  why does it start with a thus?  what does it follow from?  %Granvilliers and Benhamou~\cite{gb01} propose an algorithm that allows for tight integration of interval Newton methods and constraint propagation for achieving maximal efficiency in reducing search space.

Similar heuristics are also utilized for solution of MINLP problems. For example, the aggressive probing strategy of Nannicini et al.~\cite{nbllmw11} solves probing problems to global optimality and thus has a huge computational overhead. To avoid excessive work, Nannicini et al. propose a strategy based on support vector machines~\cite{ss01} to predict when this aggressive probing strategy is likely to succeed based on the success of propagation operations. Aggressive probing is only carried out when its chances of success are high. Couenne solves linear versions of Problem~\ref{eq:obbt} in the branch-and-bound tree in all nodes up to a depth specified by a parameter $L$. For nodes at a depth $d>L$, the strategy is applied with a probability $2^{L-d}$. A similar strategy is employed by ANTIGONE. The idea of propagation of Lagrangian variable bounds~\cite{gw13,gbmw16} can also be thought of as a means of balancing the computational effort for tightening based on solving variants of Problem~\ref{eq:obbt}. Vu et al.~\cite{vss09} show significant computational benefits  by carrying out propagation on a subset of the nodes and a partial subgraph of the DAG rather than the entire graph. Vu et al.~\cite{vsb09} present multiple strategies for combining the various reduction schemes from constraint programming and mathematical programming for constraint satisfaction problems.

\section{Computational impact of domain reduction}
\label{sec:comp}
We demonstrate the computational impact of domain reduction techniques on three widely available solvers: BARON~\cite{ts:comp:04}, Couenne~\cite{b+:09:oms} and SCIP~\cite{a09:scip}. BARON is commercial~\cite{tofbaron} and also free through the NEOS server~\cite{cmm98}.  SCIP is free for academics and commercial for all others.  Couenne is open-source and free software.  All three solvers are available under the GAMS modeling system~\cite{gams} and provide options that allow us to turn off their domain reduction algorithms.  The global MINLP solvers Antigone~\cite{mf:14:anti} and LindoGlobal~\cite{ls:09:oms} are also available under GAMS but were not used in these experiments since they do not offer facilities that turn off their presolve routines.

The test libraries used in our tests are the Global library~\cite{globallib}, Princeton library~\cite{princetonlib}, MINLP library~\cite{minlplib} and the CMU-IBM library~\cite{ibmlib}. Global and Princeton libraries consist of NLP problems, while the MINLP and CMU-IBM libraries consist of MINLP problems. While over 25\% of the Princeton library are convex and the NLP relaxations of the CMU-IBM library problems are all convex, the Global and MINLP libraries contain mostly nonconvex problems. In the sequel we present results for each library separately so as to demonstrate that the observed trends are not dominated by any particular library, convexity, or integrality properties.  Key statistics on the test sets are summarized in Table~\ref{tb:stat}.

\begin{table*}[htb]
\centering
\begin{tabular}{|l|c|c|c|c|}
\hline
Test Library & Global & Princeton & CMU-IBM & MINLP\\
\hline
Number of problems & 369 & 980 & 142 & 249 \\
\hline
Avg. no. of continuous variables & 1092 & 1355 & 369 & 346 \\
\hline
Avg. no. of binary variables & 0 & 0 & 139 & 235 \\
\hline
Avg. no. of integer variables & 0 & 0 & 0 & 24 \\
\hline
Avg. no. of constraints & 785 & 836 & 956 & 534\\
\hline
\end{tabular}
\caption{Test set statistics}
\label{tb:stat}
\end{table*}

All computational tests were run on a 64-bit Intel Xeon X5650 2.66 Ghz processor running CentOS release 7. The tests were carried out with a time limit of 500 seconds and absolute and relative optimality tolerances set to $10^{-6}$.  All solvers were run under two different settings: (1) default options and (2) default options with domain reduction turned off.  Our main objective is to compare these two settings. Comparing the relative impact of the individual reduction techniques on the performance of global solvers is beyond the scope of this work. Readers can refer to K{\i}l{\i}n{\c{c}} and Sahinidis \cite{ks14} for experiments showing the effect of different bounds reduction strategies in BARON. In the results presented below, the suffix ``nr" is used to denote a solver applied with domain reduction techniques turned off.  Comparisons are presented in terms of performance profiles generated through PAVER~\cite{mp06}.  In these profiles, a solver is considered to have solved a problem if it obtains the best solution amongst the solvers compared in the profile within a given multiple of the time taken by the fastest solver that solves a problem. CAN\_SOLVE denotes the number of problems in the library that can be solved with all the solvers compared in the profile.

\subsection{BARON}
The solver options and their values utilized for tests with BARON are summarized in Table~\ref{tb:baropt}. These turn off the various domain reduction techniques utilized in BARON. The remaining options are utilized at their default settings.
\begin{table}[htb]
\centering
\begin{tabular}{|l|c|}
\hline
Option &  Value \\
\hline
LBTTDo &  0 \\
\hline
MDo &  0 \\
\hline
OBTTDo &  0 \\
\hline
PDo &  0 \\
\hline
TDo &  0 \\
\hline
\end{tabular}
\caption{BARON options used for tests}
\label{tb:baropt}
\end{table}

\begin{figure*}[htbp]
\centering
\includegraphics[scale = 0.4]{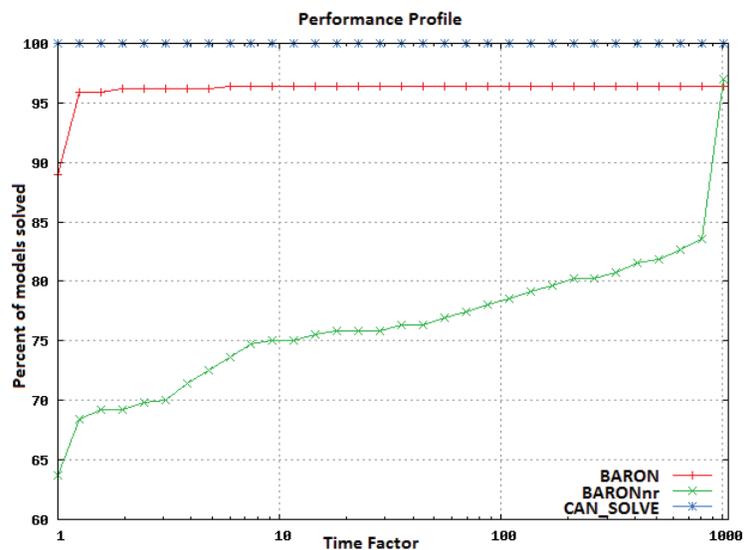}
\caption{Performance profiles for BARON on Global library}
\label{fig:baron_global}
\end{figure*}

\begin{figure*}[htbp]
\centering
\includegraphics[scale = 0.4]{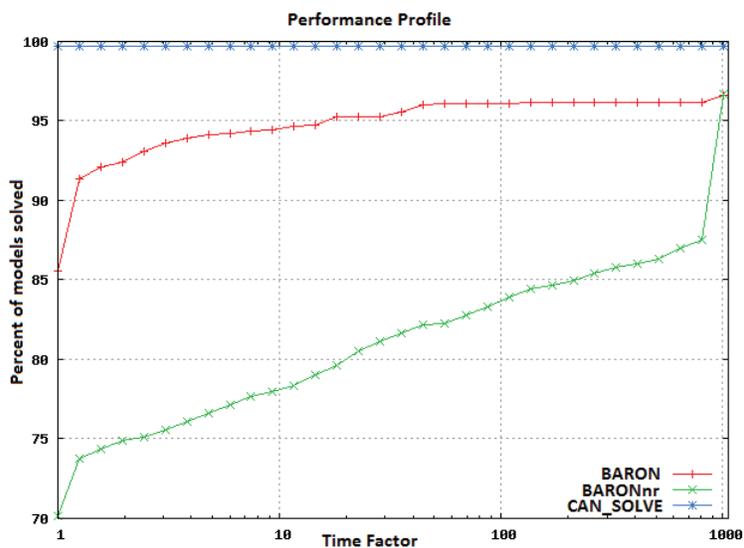}
\caption{Performance profiles for BARON on Princeton library}
\label{fig:baron_prince}
\end{figure*}

\begin{figure*}[htbp]
\centering
\includegraphics[scale = 0.4]{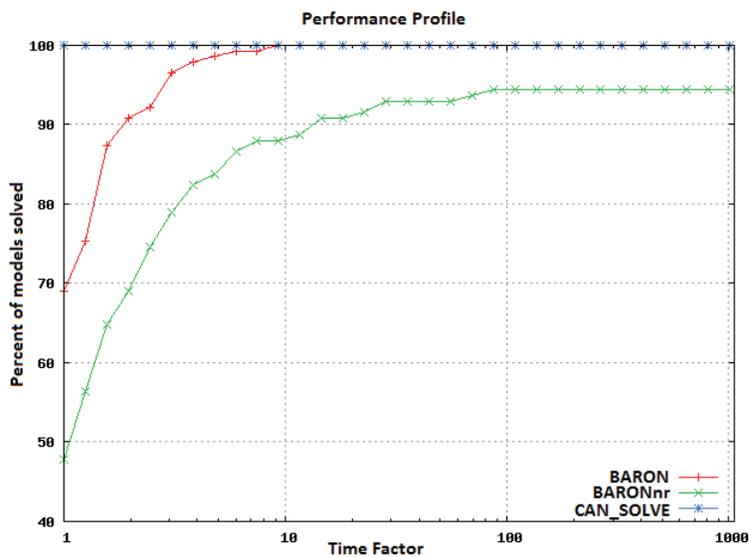}
\caption{Performance profiles for BARON on CMU-IBM library}
\label{fig:baron_ibm}
\end{figure*}

\begin{figure*}[htbp]
\centering
\includegraphics[scale = 0.4]{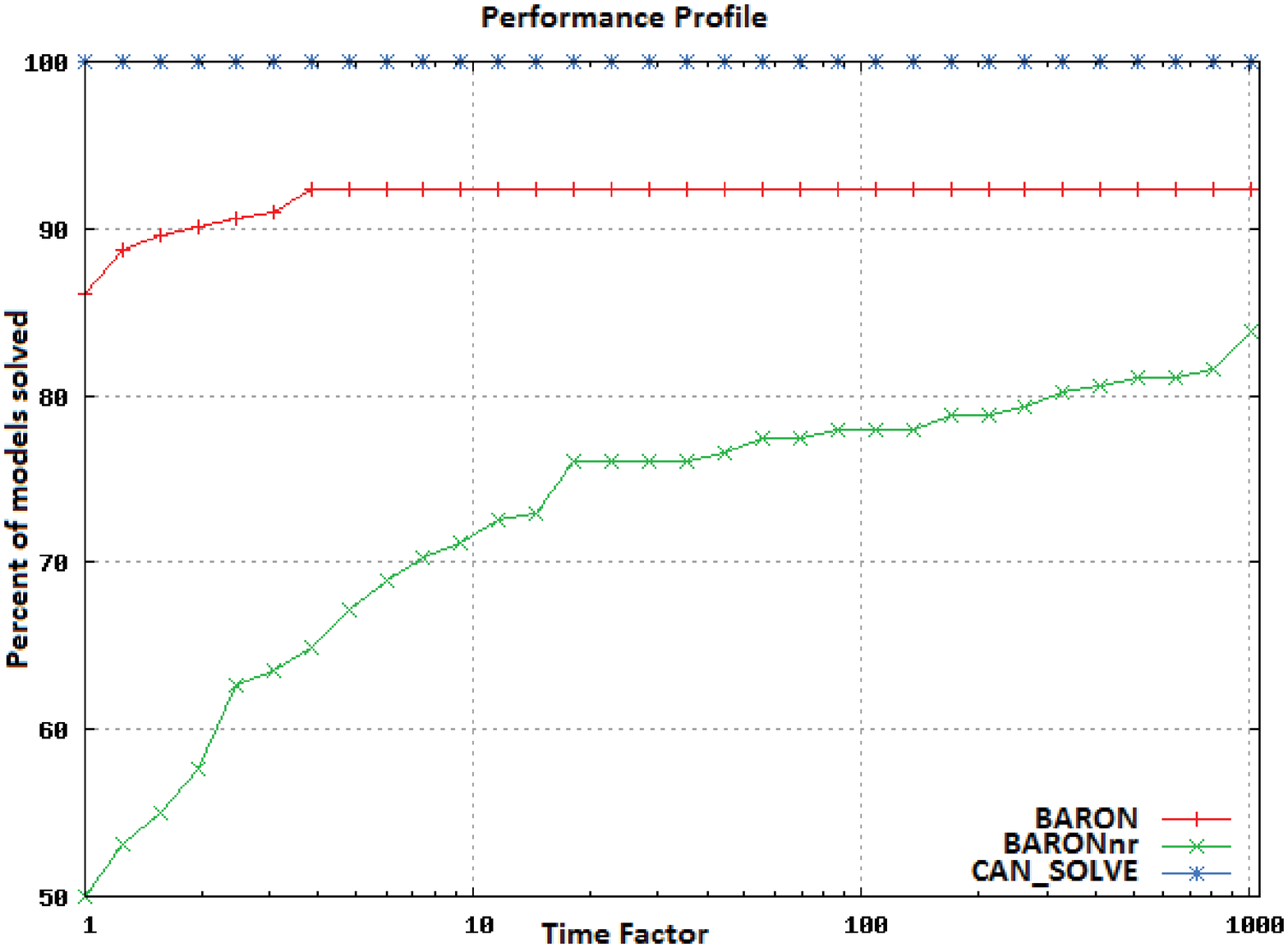}
\caption{Performance profiles for BARON on MINLP library}
\label{fig:baron_minlp}
\end{figure*}

Figures~\ref{fig:baron_global},~\ref{fig:baron_prince},~\ref{fig:baron_ibm} and~\ref{fig:baron_minlp} demonstrate the performance of BARON with and without domain reduction techniques employed on the Global, Princeton, CMU-IBM and MINLP libraries. The profiles indicate a huge deterioration in performance across all test libraries when reduction is turned off.  Interestingly, for the continuous test libraries, the performance profile of the no-reduction version of BARON ``catches up'' with that of the reduction-based version at the end of the profiles.  This simply means that, without reduction, BARON's heuristics are still able to come up with (near-)global solutions but the branch-and-bound algorithm is not able to provide sufficiently strong lower bounds in order to prove global optimality.  For the MINLP case, the no-reduction-based algorithm is not even able to find good feasible solutions.

\begin{table*}[htb]
\centering
\begin{tabular}{|l|c|c|c|c|}
\hline
Test Library & Global & Princeton & CMU-IBM & MINLP\\
\hline
\% increase in number of nodes & 1180 & 261 & 546 & 802 \\
\hline
\% increase in computational time & 67 & 70 & 75 & 47 \\
\hline
\end{tabular}
\caption{Performance deterioration for BARON when domain reduction is turned off}
\label{tb:barperf}
\end{table*}

Results in Table~\ref{tb:barperf} show that turning off domain reduction techniques leads to a huge increase in the number of nodes required by BARON. The increase in nodes is also accompanied by a significant increase in computational time across all test libraries.

\subsection{Couenne}
The solver options and their values utilized for tests with Couenne are summarized in Table~\ref{tb:couenneopt}. These turn off the various domain reduction techniques utilized in Couenne. The remaining options are utilized at their default settings.
\begin{table}[htb]
\centering
\begin{tabular}{|l|c|}
\hline
Option &  Value \\
\hline
aggressive\_fbbt   & no \\
\hline
branch\_fbbt &  no \\
\hline
feasibility\_bt & no \\
\hline
max\_fbbt\_iter  & 0 \\
\hline
optimality\_bt & no \\
\hline
redcost\_bt & no\\
\hline
fixpoint\_bt & no \\
\hline
two\_implied\_bt & no \\
\hline
\end{tabular}
\caption{Couenne options used for tests}
\label{tb:couenneopt}
\end{table}

\begin{figure*}[htbp]
\centering
\includegraphics[scale = 0.4]{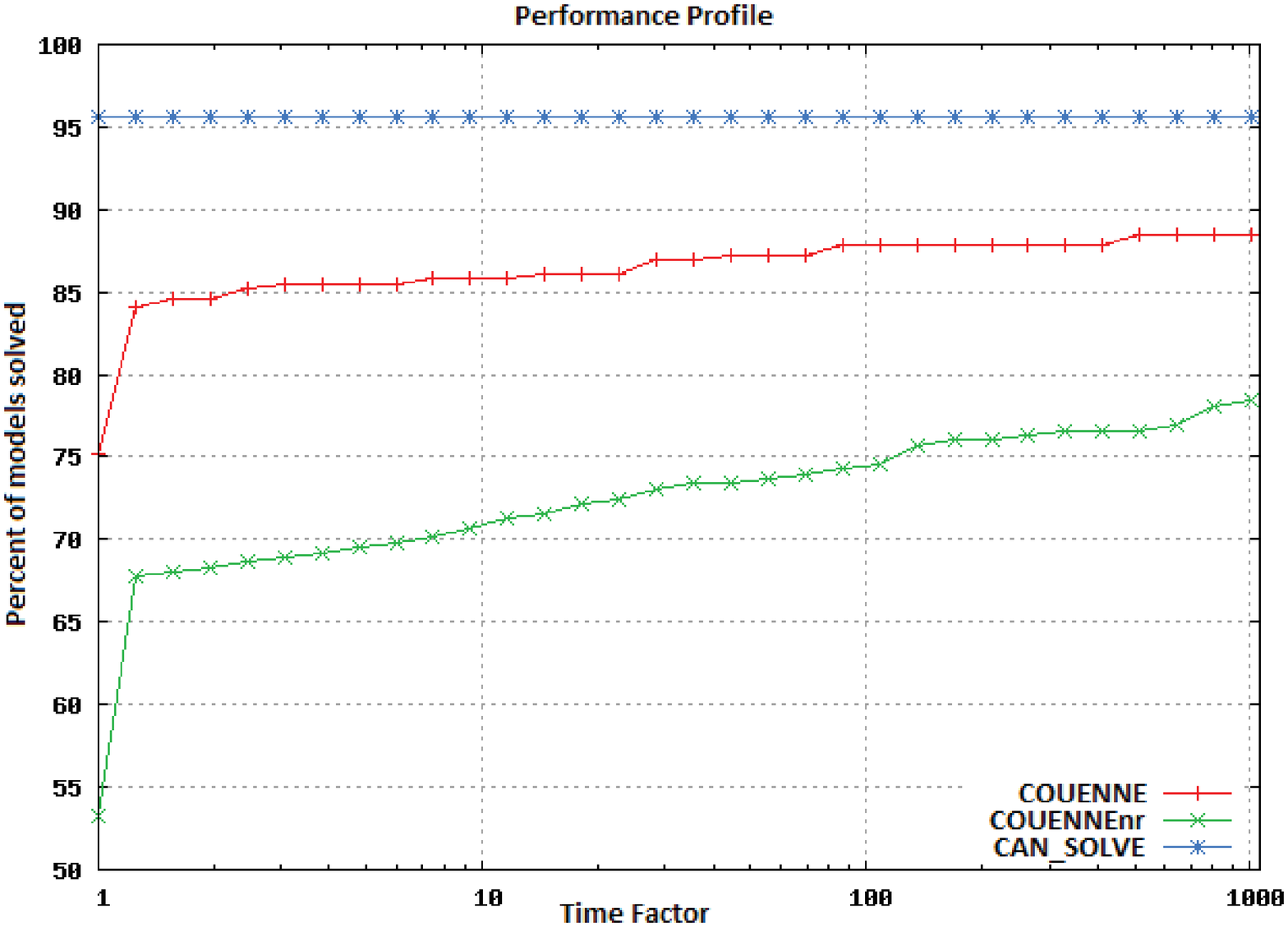}
\caption{Performance profiles for Couenne on Global library}
\label{fig:couenne_global}
\end{figure*}

\begin{figure*}[htbp]
\centering
\includegraphics[scale = 0.4]{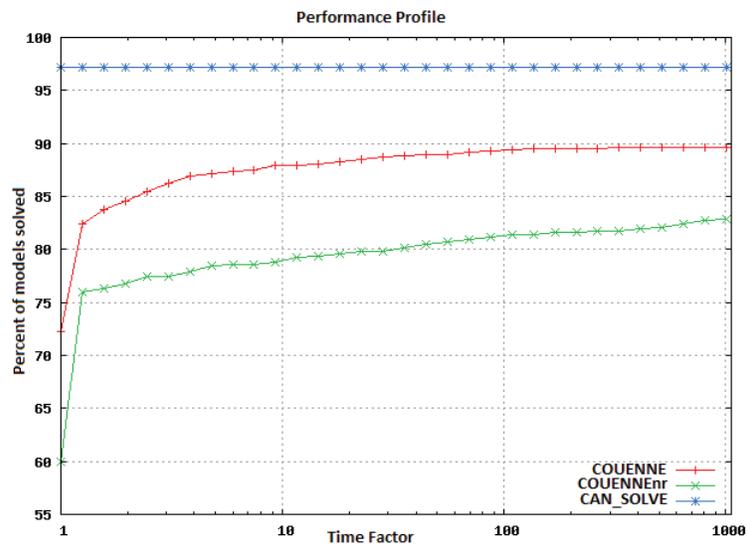}
\caption{Performance profiles for Couenne on Princeton library}
\label{fig:couenne_prince}
\end{figure*}

\begin{figure*}[htbp]
\centering
\includegraphics[scale = 0.4]{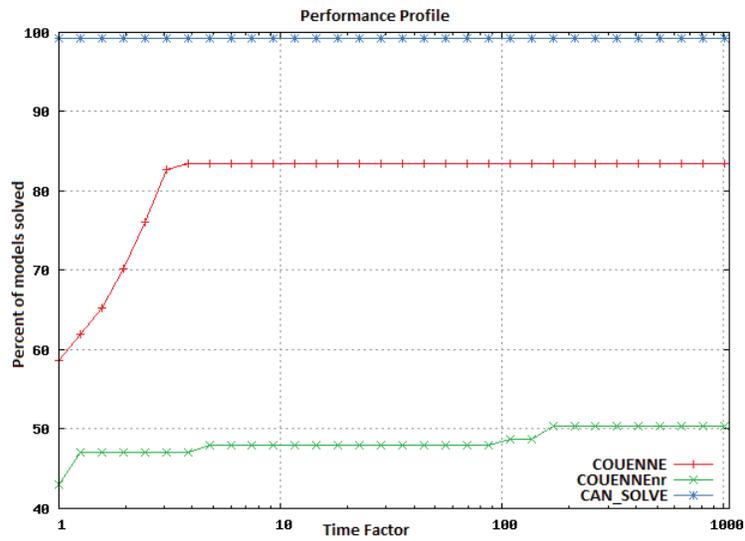}
\caption{Performance profiles for Couenne on CMU-IBM library}
\label{fig:couenne_ibm}
\end{figure*}

\begin{figure*}[htbp]
\centering
\includegraphics[scale = 0.4]{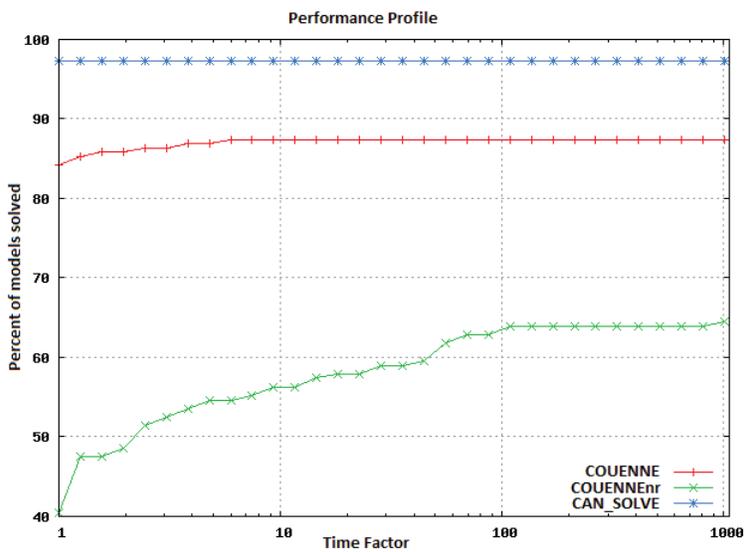}
\caption{Performance profiles for Couenne on MINLP library}
\label{fig:couenne_minlp}
\end{figure*}

Figures~\ref{fig:couenne_global},~\ref{fig:couenne_prince}, \ref{fig:couenne_ibm} and~\ref{fig:couenne_minlp} demonstrate the performance of Couenne with and without domain reduction techniques employed on the Global, Princeton, CMU-IBM and MINLP libraries. Similar to BARON, turning off domain reduction techniques has a huge impact on the performance of Couenne. Contrary to BARON, without reduction, this solver is unable to find good solutions for the NLP test libraries; as a result, the performance profiles for the no-reduction-based algorithm do not catch up with the reduction-based algorithm.  Table~\ref{tb:Couenneperf} indicates a significant increase in computational time and the number of nodes explored in the branch-and-bound search when reduction is turned off. It should be pointed out that no comparisons are possible between BARON and Couenne by looking at their respective performance profiles since these profiles depend solely on the solvers included in each figure.

\begin{table*}[htb]
\centering
\begin{tabular}{|l|c|c|c|c|}
\hline
Test Library & Global & Princeton & CMU-IBM & MINLP\\
\hline
\% increase in number of nodes & 129 & 21 & 186 & 171 \\
\hline
\% increase in computational time & 26 & 19 & 12 & 16 \\
\hline
\end{tabular}
\caption{Performance deterioration for Couenne when domain reduction is turned off}
\label{tb:Couenneperf}
\end{table*}

\subsection{SCIP}
The solver options and their values utilized for tests with SCIP are summarized in Table~\ref{tb:scipopt}. As before, the remaining options are used at their default settings.

\begin{table*}[htb]
\centering
\begin{tabular}{|l|c|c|}
\hline
Option &   Value \\
\hline
presolving/maxrounds & 0 \\
\hline
presolving/components/maxrounds & 0\\
\hline
presolving/convertinttobin/maxrounds &0\\
\hline
presolving/domcol/maxrounds & 0\\
\hline
presolving/dualagg/maxrounds & 0\\
\hline
presolving/dualinfer/maxrounds & 0\\
\hline
presolving/gateextraction/maxrounds & 0\\
\hline
presolving/implfree/maxrounds & 0\\
\hline
presolving/implics/maxrounds & 0\\
\hline
presolving/inttobinary/maxrounds & 0\\
\hline
presolving/redvub/maxrounds & 0\\
\hline
presolving/stuffing/maxrounds & 0\\
\hline
presolving/trivial/maxrounds & 0\\
\hline
presolving/tworowbnd/maxrounds & 0\\
\hline
propagating/maxrounds & 0\\
\hline
propagating/dualfix/maxprerounds & 0\\
\hline
propagating/genvbounds/maxprerounds & 0\\
\hline
propagating/obbt/freq & -1\\
\hline
propagating/obbt/maxprerounds & 0\\
\hline
propagating/obbt/tightintboundsprobing & False \\
\hline
propagating/probing/maxprerounds & 0\\
\hline
propagating/pseudoobj/maxprerounds & 0\\
\hline
propagating/redcost/maxprerounds & 0\\
\hline
propagating/rootredcost/maxprerounds & 0\\
\hline
propagating/vbounds/maxprerounds & 0\\
\hline
conflict/useprop & False \\
\hline
conflict/useinflp & False \\
\hline
conflict/usepseudo & False \\
\hline
heuristics/bound/maxproprounds & 0\\
\hline
heuristics/clique/maxproprounds & 0 \\
\hline
heuristics/randrounding/maxproprounds & 0 \\
\hline
heuristics/shiftandpropagate/freq & -1 \\
\hline
heuristics/shifting/freq & -1 \\
\hline
heuristics/vbounds/maxproprounds & 0 \\
\hline
misc/allowdualreds & False \\
\hline
misc/allowobjprop & False \\
\hline
\end{tabular}
\caption{SCIP options used for tests}
\label{tb:scipopt}
\end{table*}

\begin{figure*}[htbp]
\centering
\includegraphics[scale = 0.4]{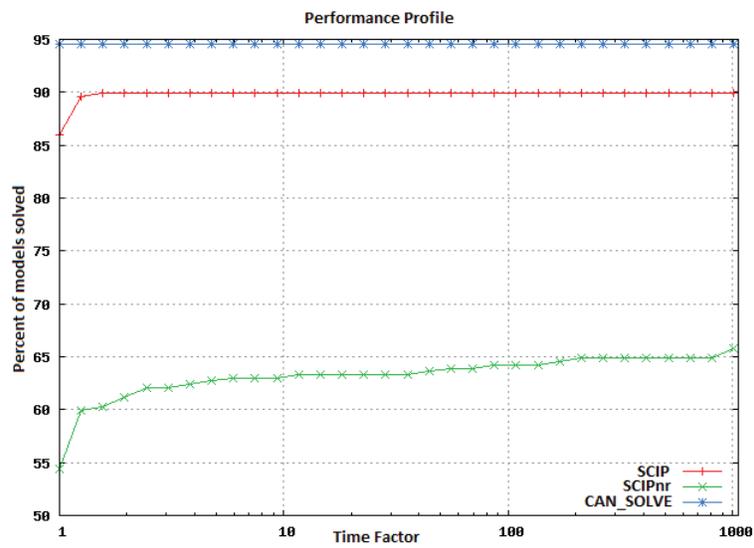}
\caption{Performance profiles for SCIP on Global library}
\label{fig:scip_global}
\end{figure*}

\begin{figure*}[htbp]
\centering
\includegraphics[scale = 0.4]{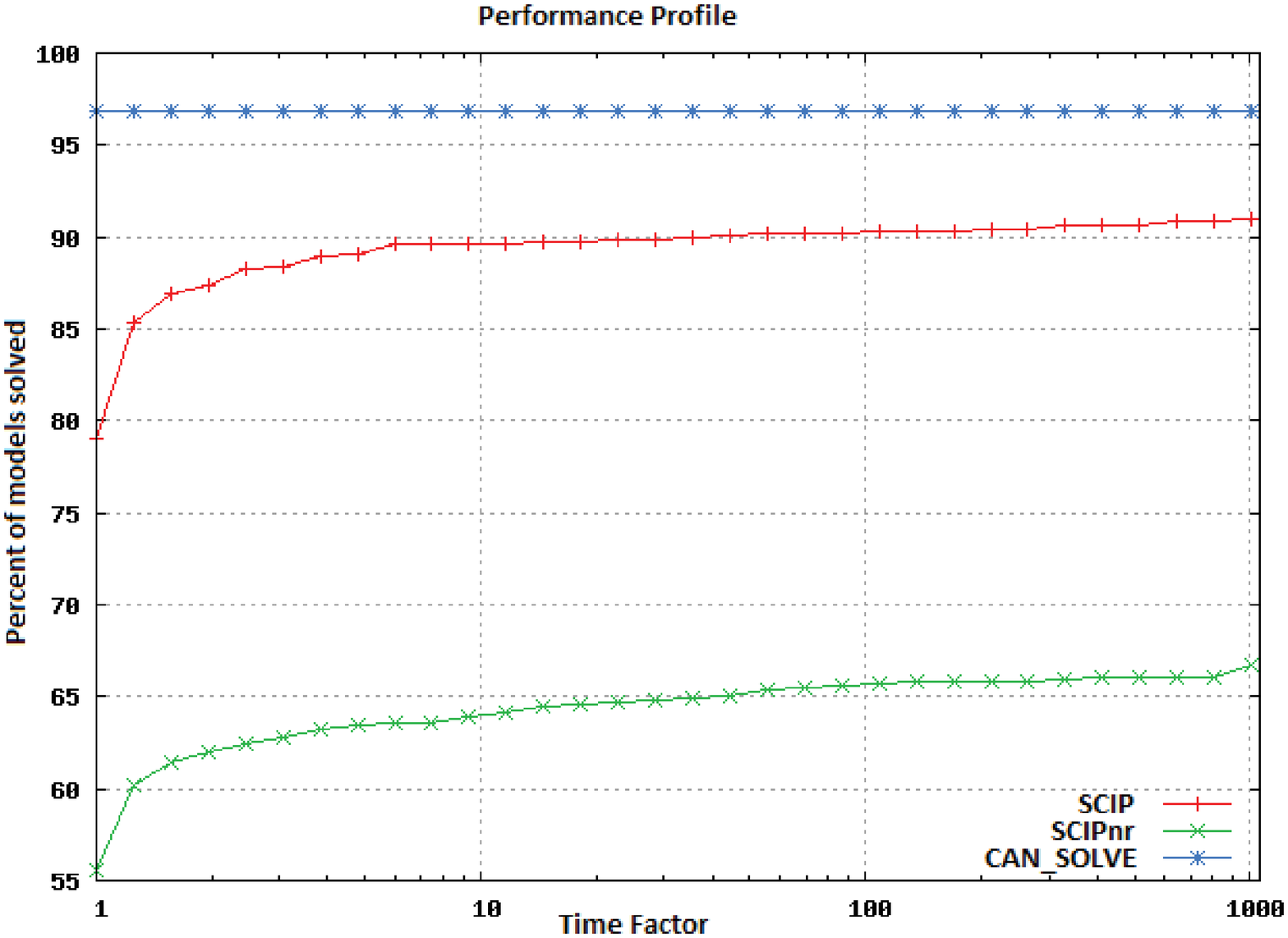}
\caption{Performance profiles for SCIP on Princeton library}
\label{fig:scip_prince}
\end{figure*}

\begin{figure*}[htbp]
\centering
\includegraphics[scale = 0.4]{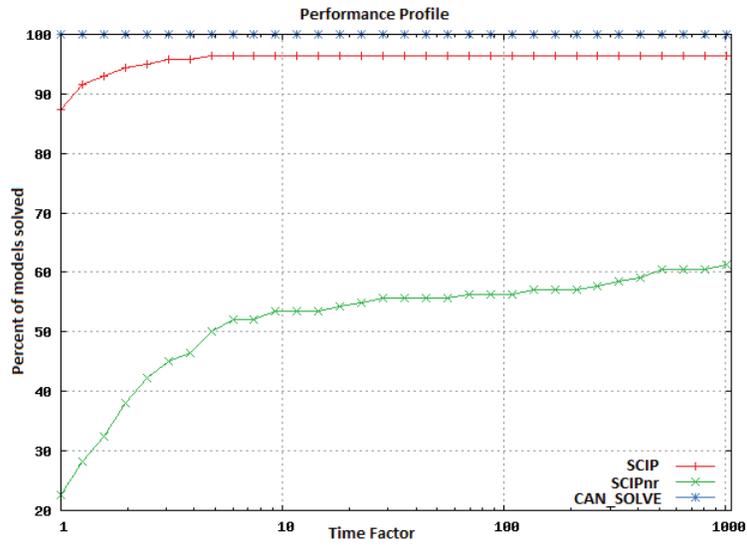}
\caption{Performance profiles for SCIP on CMU-IBM library}
\label{fig:scip_ibm}
\end{figure*}

\begin{figure*}[htbp]
\centering
\includegraphics[scale = 0.4]{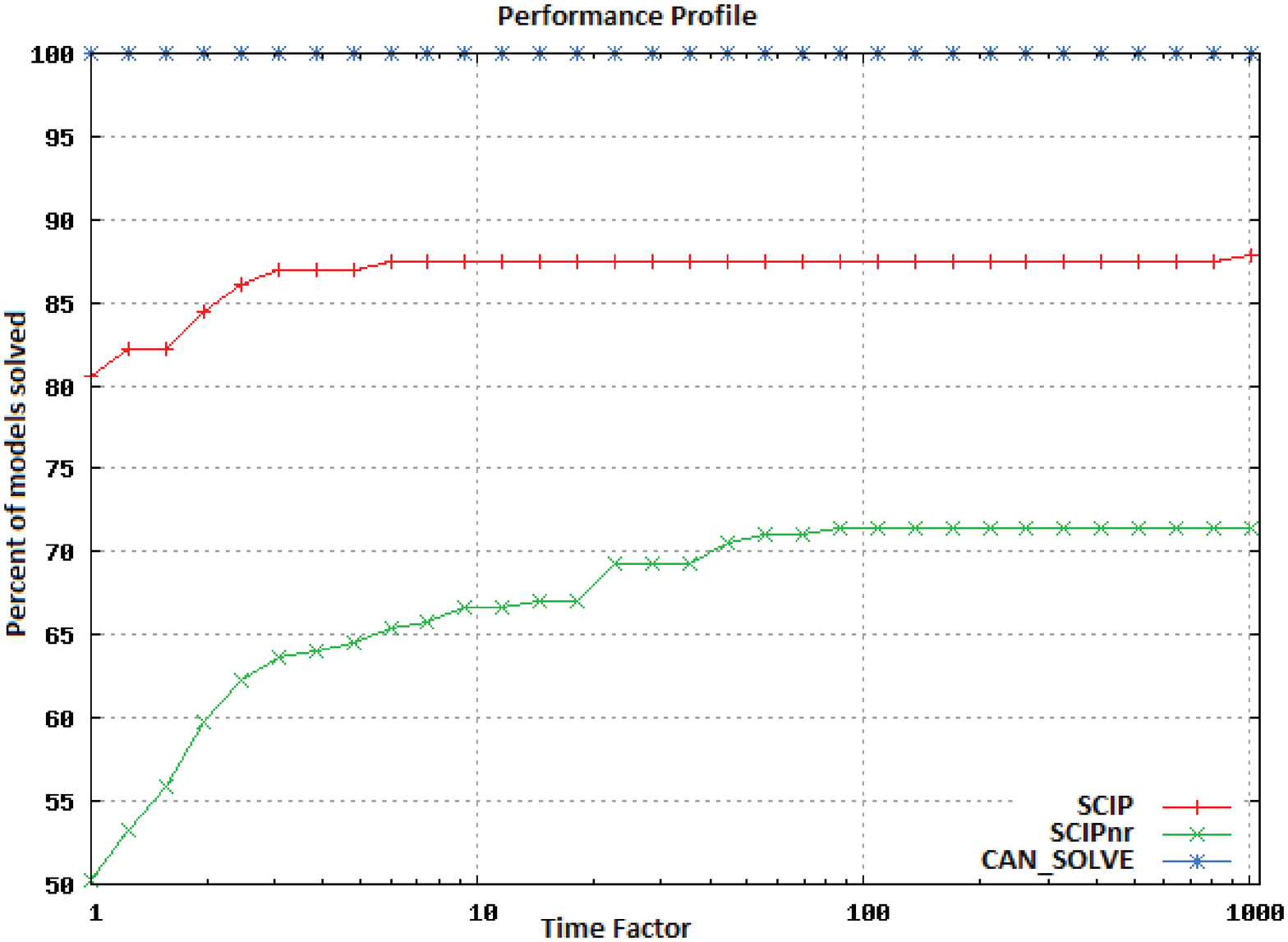}
\caption{Performance profiles for SCIP on MINLP library}
\label{fig:scip_minlp}
\end{figure*}

Figures~\ref{fig:scip_global},~\ref{fig:scip_prince}, \ref{fig:scip_ibm} and~\ref{fig:scip_minlp} demonstrate the huge impact of turning off domain reductions with SCIP on the Global, Princeton, CMU-IBM and MINLP libraries. Similar to Couenne, this solver is also not able of finding good solutions for continuous problems when reduction is turned off.  Results in Table~\ref{tb:scipperf} indicate a significant deterioration in performance for SCIP as other solvers.

\begin{table*}[htb]
\centering
\begin{tabular}{|l|c|c|c|c|}
\hline
Test Library & Global & Princeton & CMU-IBM & MINLP\\
\hline
\% increase in number of nodes & 174 & 152 & 417 & 56 \\
\hline
\% increase in computational time & 24 & 33 & 141 & 18 \\
\hline
\end{tabular}
\caption{Performance deterioration for SCIP when domain reduction is turned off}
\label{tb:scipperf}
\end{table*}

\subsection{Relative solver performance}

\begin{figure*}[htbp]
\centering
\includegraphics[scale = 0.4]{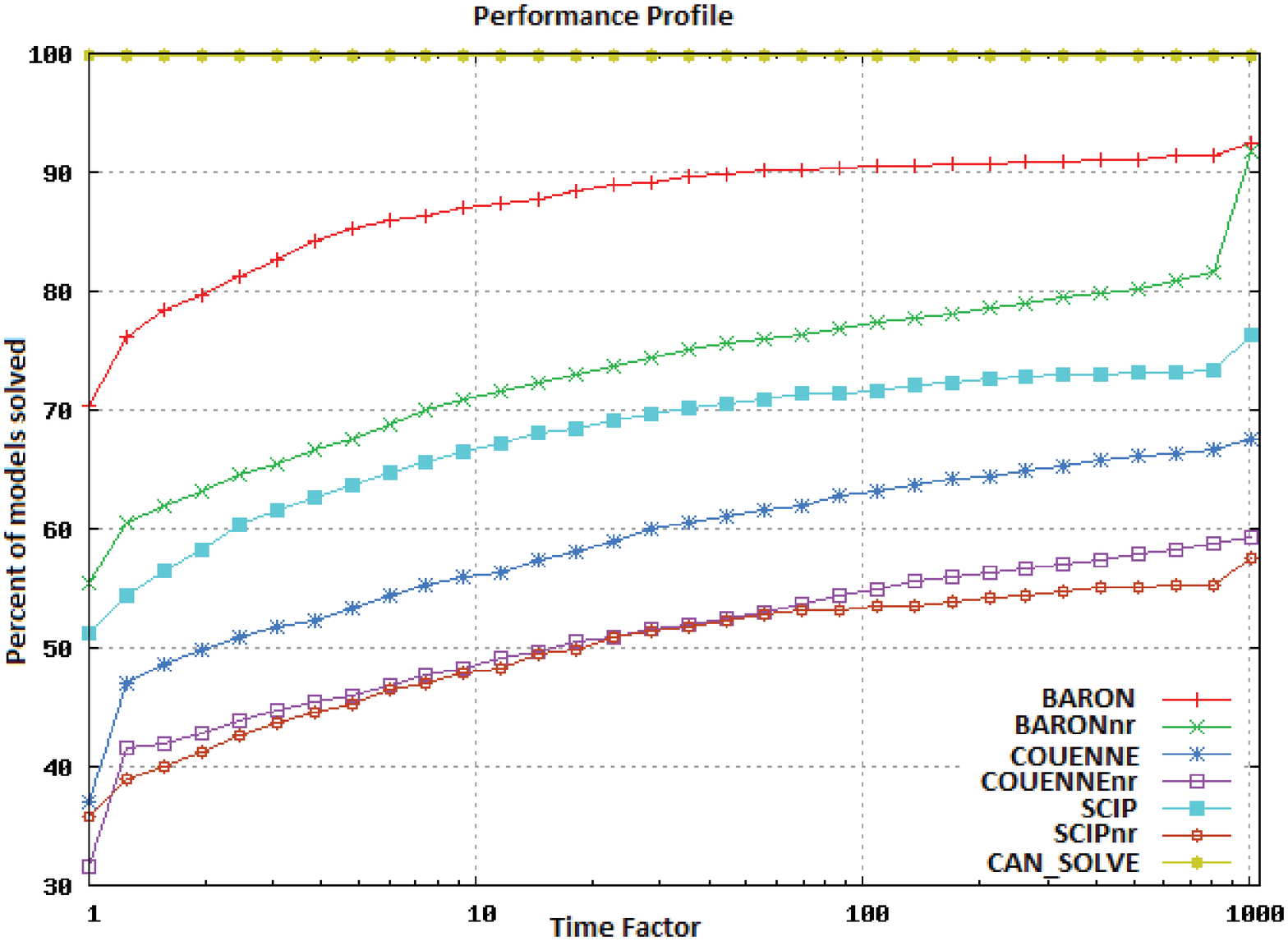}
\caption{Performance profiles for all solvers on all libraries combined}
\label{fig:combined}
\end{figure*}

Figure~\ref{fig:combined} describes the performance of BARON, Couenne and SCIP on all test libraries aggregated together. Even without reduction, BARON dominates over the other two solvers, suggesting that this solver has an edge over the other two solvers in terms of its technology for relaxation construction, branching schemes, and primal feasibility heuristics.  However, the impact of domain reduction techniques is clear and rather substantial on all three solvers.  Interestingly, the relative order of SCIP and Couenne changes when reduction is turned off, suggesting that SCIP is relying much more on domain reduction than Couenne does.  Equivalently, SCIP enjoys a substantial advantage over Couenne thanks to its implementation of a more extensive set of reduction techniques.

\section{Conclusions}
\label{conclusions}

We have presented a review of the various domain reduction techniques proposed in literature for the purpose of global NLP and MINLP optimization. These techniques vary in complexity including simple ones like propagation to more computationally intensive ones that involve full solution of optimization subproblems. Application of some of the more complex techniques requires the use of smart heuristics to ensure that they are utilized only when domain reduction is likely. We have also presented computational results with BARON, SCIP and Couenne on publicly available test libraries. The results show that domain reduction techniques have a significant impact on the performance of these solvers. Incorporation of domain reduction within branch-and-bound leads to huge reductions in computational time and number of nodes required for solution. Future research in this area should focus on the development of domain reduction techniques based on sets of constraints (i.e., global filtering methods) for broad classes of structured problems but also for important science and engineering problems, such as pooling problems~\cite{rs17}, protein folding~\cite{n97-pro}, and network design problems~\cite{da+:15,rb:15}.

\bibliographystyle{spmpsci}      % mathematics and physical sciences
%\bibliography{ypp,reference}   % name your BibTeX data base

\end{singlespacing}

\end{document}